# Focal and Connectomic Mapping of Transiently Disrupted Brain Function

Michael S. Elmalem[1,2,3], Hanna Moody[1], James K. Ruffle[1,2], Michel Thiebaut de Schotten[4,5], Patrick Haggard[6], Beate Diehl[1,2], Parashkev Nachev[1,2], and Ashwani Jha[1,2]

**Affiliation(s):**

[1] UCL Queen Square Institute of Neurology, London, United Kingdom
   Department of Brain Repair and Rehabilitation
   High Dimensional Neurology Group
[2] National Hospital for Neurology and Neurosurgery, London, United Kingdom
[3] Max Planck Institute for Human Cognitive and Brain Sciences, Leipzig, Germany
[4] Groupe d'Imagerie Neurofonctionnelle, Institut des Maladies Neurodégénérative, University of Bordeaux, Bordeaux, France
[5] Brain Connectivity and Behaviour Laboratory, Sorbonne Universities, Paris, France
[6] UCL Institute of Cognitive Neuroscience, London, United Kingdom

**Corresponding Authors:**

**Parashkev Nachev & Ashwani Jha**
UCL Queen Square Institute of Neurology, London, United Kingdom
Department of Brain Repair and Rehabilitation
High Dimensional Neurology Group
Email addresses: p.nachev@ucl.ac.uk | ashwani.jha@ucl.ac.uk

## Abstract

The distributed nature of the neural substrate, and the difficulty of establishing necessity from correlative data, combine to render the mapping of brain function a far harder task than it seems. Methods capable of combining connective anatomical information with focal disruption of function are needed to disambiguate local from global neural dependence, and critical from merely coincidental activity. Here we present a comprehensive framework for focal and connective spatial inference based on sparse disruptive data, and demonstrate its application in the context of transient direct electrical stimulation of the human medial frontal wall during the pre-surgical evaluation of patients with focal epilepsy. Our framework formalizes voxel-wise mass-univariate inference on sparsely sampled data within the statistical parametric mapping framework, encompassing the analysis of distributed maps defined by any criterion of connectivity. Applied to the medial frontal wall, this *transient dysconnectome* approach reveals marked discrepancies between local and distributed associations of major categories of motor and sensory behaviour, revealing differentiation by remote connectivity to which purely local analysis is blind. Our framework enables disruptive mapping of the human brain based on sparsely sampled data with minimal spatial assumptions, good statistical efficiency, flexible model formulation, and explicit comparison of local and distributed effects.

# 1. Introduction

Three decades into the human brain mapping revolution ushered by functional magnetic resonance imaging (MRI), large swathes of the neural landscape remain shrouded in darkness. Two cardinal aspects of the task are increasingly recognised to inhibit progress: the *distributed*, connective nature of the neural substrate (Alivisatos et al., 2012; Catani et al., 2012), and the difficulty of establishing *necessity* from predominantly correlative data (Adolphs, 2016; Rorden & Karnath, 2004). Each aspect on its own presents formidable difficulties: characterising distributed substrates requires explicit modelling of remote interactions intractable without large-scale data and mathematical models embrittled by their complexity; establishing necessity requires disruptive evidence typically obtained naturally, through the behavioural consequences of uncontrolled focal pathological lesions confounded by their incidental—and heterogeneously distributed—characteristics (Mah et al., 2014; Xu et al., 2018). Combined, these difficulties are reciprocally amplified: data of sufficient scale and quality to support complex models is especially hard to acquire in the clinical domain, and distributed patterns of pathological damage become entangled with the comparably distributed underlying patterns of neural dependence (with rare exceptions (Jha et al., 2020)). Yet it is precisely distributed substrates that are most in need of disruptive evidence, for the plurality of neural support makes inferences from correlative data all the harder.

Methodological innovation at the intersection of connective and disruptive mapping of brain function is therefore urgently needed, with close attention not only to the practicalities of scaling current techniques, but also to diminishing the need for data volumes that will always be hard to achieve. Here we elaborate conceptually, implement technically, and demonstrate empirically, a simple, robust, and efficient approach to connective disruptive mapping of human brain function in the clinical context of direct cortical electrical stimulation (DCS).

Theoretically, the ideal approach is to register the functional consequences of transient disruption applied at single point loci, individually and in combination, across the entire brain. DCS, commonly employed as a localising clinical tool in patients undergoing evaluation for resective surgery of (typically epileptogenic) lesions, approximates this ideal arguably closer than any other available tool. Focal, transient disruption can thereby be achieved, enabling causally more robust examination of the relationship between a well-defined neural substrate and an observed, or reported, behavioural outcome (Desmurget & Sirigu, 2015). Though clinical imperatives inevitably constrain the choice of locations and sampling density, the ability to evaluate multiple loci in each patient, dynamically, yields higher volumes of informative data than the bare number of surveyed patients suggests. The approach has already been extensively used to derive maps of functional dependence (Roux et al., 2003) in surgical settings (Sarubbo et al., 2015, 2020), including connectivity (Corrivetti et al., 2019), but outside a formal framework that allows both focal and connective effects to be robustly quantified without dependence on predefined regions of interest.

For all its theoretical power, the use of DCS for spatial inference is complicated by its sparsity. Although it is common practice to evaluate multiple loci in each individual patient—and grid electrodes may offer locally dense coverage—comprehensive sampling at high resolution across the brain is infeasible. The traditional solution is to adopt an *a priori* region of interest (ROI) parcellation, and report behaviour averaged across each sampled region (Kim et al. 2010, Goldstein et al., 1999; Fischl et al., 2004). Informing regional parcellations by richer representations of neural similarity such as histology, functional, and structural connectivity (e.g., Glasser et al., 2016) increases our confidence in their fidelity, but only as far as these characteristics may reasonably be taken as indicators of functional homology: a question that can be definitively settled only by disruptive techniques itself. Moreover, this approach to anatomical inference has six defects. First, it assumes that the constituents of each ROI are homogeneous and interchangeable, a simplistically modular, "Lego" vision of the brain not sustainable on close examination (Amunts et al., 1999). Second, it assumes that the ROI allocation of a given stimulated locus is both certain and invariant to its distance from the ROI boundary, neither of which is plausible. Third, the resultant inference presupposes the topology it is supposed to reveal, for it is expressed in a parcellation defined before the data is even acquired. Fourth, no regional difference will register where the true functional boundary is orthogonal to the a priori one. Fifth, where a functional pattern exhibits a finer anatomical organisation than the *a priori* parcellation, it will be invisible through it. Sixth, both continuous and discrete spatial variations in function will appear equally abrupt.

These defects have motivated us to develop a different approach, analogous to meta-analytic mapping (Eickhoff et al., 2009, 2012), that enables inference to the spatial characteristics of sparsely sampled critical areas without any prior assumptions on their structure beyond a reasonable degree of local smoothness (Trevisi et al., 2018). Consider in illustration a target anatomical domain—the dorsal medial wall, represented in 2D for simplicity—where a set of N discrete loci registered on a common grid are associated with two different deficits (Figure 1). Though the data is in anatomical register, we cannot perform a statistical test at every point on the grid, for no point is sufficiently sampled. The conventional solution is to aggregate the observed deficits within pre-defined ROIs, and report statistics in regional terms (Figure 1A). Framed as count regression, for example, the task is to predict the number of instances observed within an ROI (the dependent variable) given the behavioural parameters (the independent variables), iterating across ROIs. This yields a map structured by the chosen parcellation, sensitive to the correspondence between the parcellation and the underlying functional substrate. Where the two correspond poorly (Figure 1B) the inference will be distorted or fail altogether.

Our alternative approach is to transform each sparse location from a single point to a dense spatial distribution, thereby enabling point-wise mass univariate inference on a regular grid. In the simplest, *focal* form, this is achieved by convolving each point with a Gaussian of predetermined width, assuming that each sampled point is drawn from an underlying spatial distribution that a random Gaussian field can approximate. The locus of disruption is thus represented not as a point but as a continuous spatial distribution whose density gracefully captures the uncertainty of each disrupted location across the entire anatomical domain (Figure 1C). Now framed as linear regression, the task is to predict the density observed within each voxel (the dependent variable) given the behavioural parameters (the independent variables), iterating across voxels in mass-univariate fashion. Though it may seem counterintuitive to designate location density as the dependent variable and disrupted behaviour as an independent variable, the choice here is motivated by the task of spatial inference in the context of mass-univariate analysis, and is aligned with established practice in other domains, such as voxel-based morphometry and functional imaging. For example, in voxel-based morphometry, the dependent variable is tissue concentration, and behaviour is amongst the independent variables.

In the more complex, *connective* form, the transformation is achieved by a probabilistic projection of the distributed connectivity of each locus, incorporating not just the uncertainty but also the network distribution of the disruption (Foulon et al., 2018). In both cases, this transformation allows us to use the well-established, principled approach to mass univariate spatial analysis embodied in the statistical parametric mapping (SPM) platform (Friston et al., 1994).

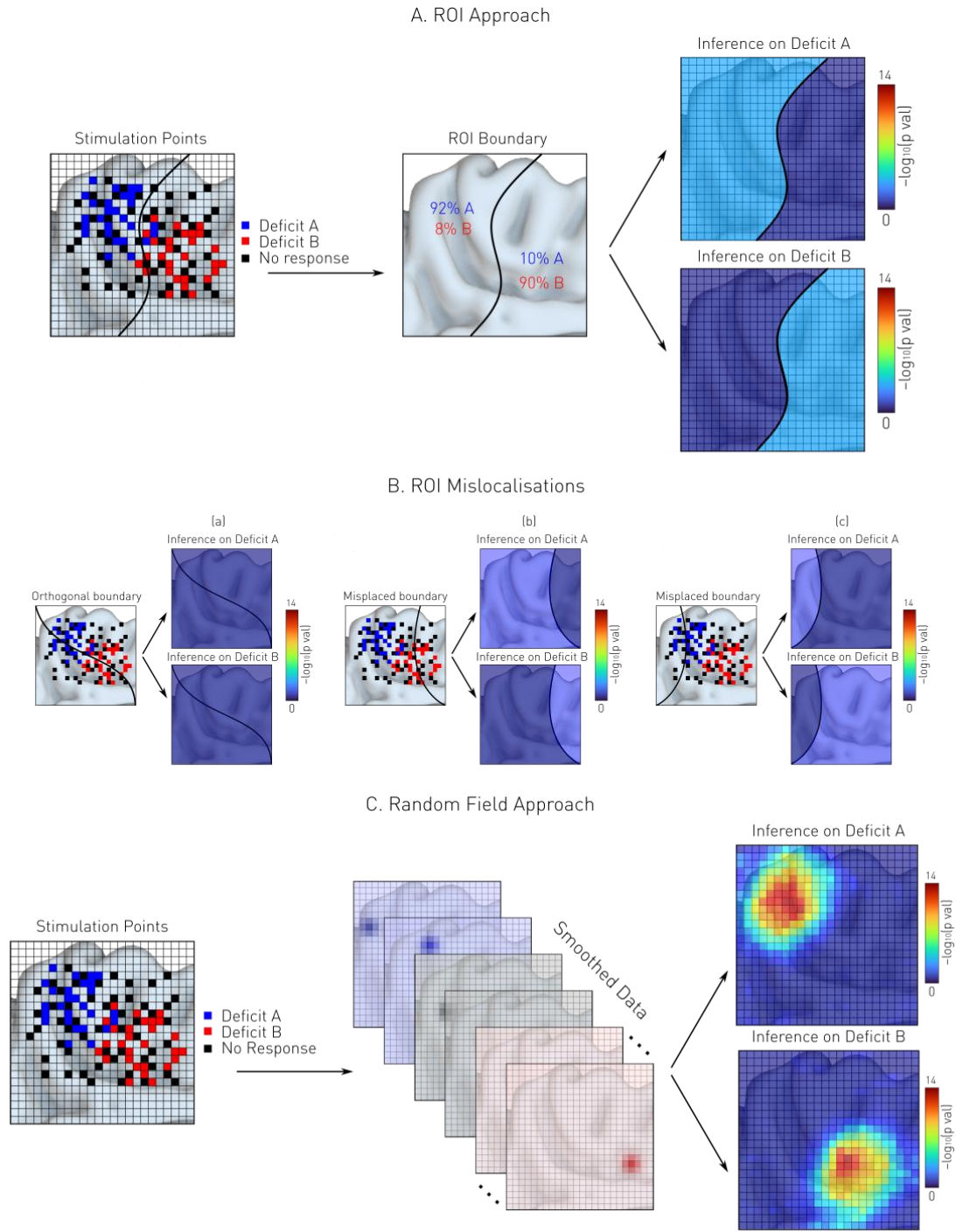

**Figure 1. ROI-based vs Random Field-based Spatial Inference. A**. *ROI-based spatial inference*. Simulated stimulation points resulting in two hypothetical deficits – A (blue) and B (red) – are counted across pre-defined ROIs, here represented in 2D space for simplicity. A statistical test is then performed on the counts to infer the spatial distribution of function in terms determined by the ROI boundary. **B**. *Parcellation-induced mislocalisation*. When there is a poor correspondence between the parcellation and the underlying functional substrate, the inference either fails completely (left) or is distorted (middle and right). **C**. *Random field-based spatial inference*. Here each stimulation point is convolved with a predefined Gaussian kernel, so that each location is now supported across the entire domain, enabling the application of voxel-wise inference on a regular grid. A statistical test is then performed at each voxel to retrieve the substrates associated with the observed deficits A and B. The colourmap is the negative decimal log of each $p$ value. Thresholding following multiple comparisons correction is not shown here for simplicity.

To demonstrate the practical application of our approach, we investigate the focal and connective organisation of the medial frontal wall in the context of sensorimotor behaviours. Surveying the literature on the medial frontal cortex reveals an inferential landscape dominated by correlative studies (Filevich et al., 2012; Lüders et al., 1995; Nachev et al., 2008; Penfield, 1950; Zilles et al., 1995) disclosing a remarkable diversity of cognitive and behavioural associations, many of them conflicting. Disruptive studies are rare here for three reasons. First, stroke—the commonest source of focal lesion data—rarely involves the medial wall, and when it does, the size and morphology of the injury limit its spatial resolving power (Mah et al., 2014). Second, natural lesions are never truly local: they enclose larger areas of tissue than are plausibly functionally uniform, resulting in confounding from collateral damage that is hard to remove (e.g., DeMarco & Turkeltaub, 2020). Third, non-invasive disruptive methods such as repetitive transcranial magnetic stimulation are either limited to the dorsal surface, or in reaching deeper presuppose its confounding collateral disruption along the way (Desmurget & Sirigu, 2015). Inferences from pathological forms of focal injury are in any event complicated by plasticity and reorganisation over time (Nudo & Friel, 1999), limiting generalisability.

Here we re-examine a previously reported set of thirty-seven patients undergoing direct electrical cortical stimulation across the medial wall in the context of clinical evaluation for surgical treatment of non-lesional epilepsy (Trevisi et al. 2018). The proximity of critical medial motor areas and the propensity for seizures to involve them justifies dense sampling of the area, achieved either by placing surface electrode grids or with multiple depth electrodes. We adapt the approach to focal voxel-wise inference pioneered by meta-analytic mapping (Eickhoff et al., 2012; Eickhoff et al., 2009) presented in our first report (Trevisi et al., 2018), reformulating it within SPM's statistical framework, and extend it to disruptive connective analysis (Boes et al., 2015; Salvalaggio et al., 2020) of a transient kind, introducing the notion of *transient dysconnectome mapping*. We speak of a 'connectome' because the inferred maps capture distributed anatomical relations defined by any chosen index of connectivity, and we prefix the term with '*dys*' (rather than the usual 'dis') because a transient intervention typically does not disconnect a network but renders it dysfunctional. Our reformulation enhances the statistical efficiency and sensitivity of the core approach, and its extension enables us to compare focal and distributed effects, not just along the medial wall, but across remote brain regions interconnected with it, establishing a comprehensive platform for transient disruptive mapping of the human brain.

## 2. Materials & methods

Published data from Trevisi et al. (2018) were reanalysed for this study. We performed two sets of analyses. The first to derive cortical maps of focal regions critical for specific categories of sensorimotor behaviours, the second to extend these maps to connected regions across the brain within the same inferential framework.

*2.1 Participants*

Retrospective data from 147 consecutive drug-resistant focal epilepsy patients undergoing intracranial recording between January 2008 and June 2015 at the National Hospital for Neurology and Neurosurgery, London as part of clinical assessment prior to epilepsy surgery were screened. The study was approved by the hospital as a retrospective evaluation of routine clinical practice.

Thirty-seven patients (28 males, 9 females, aged 19–68 years, mean = 33.86 years, SD = 11.11 years) were identified to have at least one stimulation in the medial frontal region, spatially confined to the medial wall area dorsal to the corpus callosum and rostral to the caudal bank of the marginal sulcus. Five patients had minor lesions near the supplementary motor area (SMA), and one had a lesioned SMA/paracentral lobule as evident on MRI. Table 1 summarises the patients' demographic and clinical characteristics [based on Trevisi et al. (2018)]. Further details are provided in Supplementary Table T1.

| | |
|---|---|
| **Demographics** | |
| Female (%) | 9 (24) |
| Male (%) | 28 (76) |
| Mean age, years (SD) | 33.86 (11.11) |
| **Clinical** | |
| Mean age of epilepsy onset, years (SD) | 11.43 (9.01) |
| Mean duration of epilepsy, years (SD) | 22.43 (10.05) |
| **Type of study** | |
| Grid electrodes (%) | 6 (16) |
| Grid & depth electrodes (%) | 12 (33) |
| SEEG (%) | 19 (51) |
| **Side of study** | |
| Dominant (%) | 18 (49) |
| Non-dominant (%) | 14 (38) |
| Bilateral (%) | 5 (13) |
| **Epileptogenic zone involvement** | |
| Frontal lobe (%) | 29 (78) |
| Medial frontal wall (%) | 19 (51) |
| **Abnormal MR imaging** | |
| Frontal lobe (%) | 10 (27) |
| On or near medial wall (%) | 5 (13) |
| **Medial wall resection done or planned** | 14 (38) |

**Table 1. Demographic and clinical characteristics.** Summary of demographic and clinical characteristics of the included patients with stimulations in the medial frontal cortex, based on Trevisi et al. (2018). Site of study is given relative to the language dominant hemisphere.

2.2 *Direct cortical stimulation procedures*

As previously described (Trevisi et al., 2008), depth electrodes were implanted in 19 patients using a frameless stereoelectroencephalography (SEEG) technique (Nowell et al., 2014). In the remaining 18 patients, craniotomy was performed for the placement of strips and/or grids with or without freehand insertion of additional depth electrodes. In 14 patients, intracranial recording was performed in the right hemisphere, in 18 patients in the left hemisphere, and in 5 patients bilaterally. Hemispheric dominance for language was inferred from fMRI data, not re-examined here. In 19 (51%) cases, the recording was in the dominant hemisphere, whereas in 18 (49%) cases, the electrodes were in the nondominant hemisphere. Bilateral language dominance was noted in five patients. The location of the electrodes was confirmed for all patients by post-implantation CT studies.

A clinical epileptologist and a physiologist performed one or more sessions of DCS during simultaneous video-EEG recording. Stimulations were typically performed after ictal recordings when patients were back on their baseline antiepileptic medication. Bipolar or monopolar stimulation trains were delivered with biphasic rectangular pulses of AC-current at 50 Hz, with a pulse width of 500 μs and a maximum duration of 5 seconds. The intensity was gradually increased from 0.5 to 7 mA in increments of 0.5-1 mA until the occurrence of a clinical sign or until after-discharges were detected on EEG monitoring (Kovac et al., 2014; Trevisi et al., 2018). Full details on the electrical stimulation intensities for each of the behavioural categories are provided in Trevisi et al., 2018. Stimulations accompanied a stereotyped set of test actions—rest, Barré and/or Mingazzini test, repeated movements of the upper and lower limbs, and during counting, reading, or repetitive monosyllabic verbalisation—as described in detail elsewhere (Trevisi et al., 2018).

2.3 *Behavioural analysis*

Three clinicians classified the observed behavioural responses as 'positive motor', 'negative motor' or 'speech disturbances'. Positive motor responses included involuntary, typically tonic or clonic, movements of the eye, head, limb or trunk. Negative motor responses included slowed or inhibited movement relative to experimentally specified movements, such as the inability to maintain prescribed postures. Speech disturbances included speech arrest, alteration in rhythm, involuntary speech, and hesitation. Live and post-hoc classifications (using video and audio telemetry recordings) were made by stimulating and attending clinicians. A contact was deemed silent if no response was obtained at the maximum stimulation intensity of 7 mA ('no response'). Responses after a seizure or after-discharges were excluded from the analysis.

Patients were also asked to report any evoked somatosensory responses such as cutaneous paraesthesias (tingling, touch, heat, and pain). Responses that were neither sensory nor motor, such as a reported urge to move or speak, or reported perception of motion without observed movement, were elicited only three times across the entire dataset and were therefore not modelled. Note that since each locus was evaluated with multiple tasks, more than one class of response may be associated with it: the classification is not anatomically exclusive.

2.4 *Imaging data acquisition and processing*

Preoperative structural T1-weighted imaging with an isotropic resolution of ~1 mm was acquired on a 3T magnetic resonance imaging (MRI) scanner. After implantation, non-contrasted structural whole-head CT scans with a resolution of 0.43×0.43×1.2 mm (SOMATA Definition 128-slice, Siemens Healthcare GmbH, Erlangen, Germany) were obtained to confirm the location of electrode contacts. All image processing was performed using SPM12 (http://www.fil.ion.ucl.ac.uk/spm/).

To facilitate group analysis, electrode locations were manually extracted from the CT and non-linearly transformed into the Montreal Neurological Institute (MNI) space template as described in Jha et al. (2016). In brief, for each patient, a rigid body co-registration to the standard SPM12 tissue probability map was performed for both preoperative T1-weighted MRI and postoperative CT, based on normalised mutual information with adjustment by Procrustes analysis, weighted by white and grey matter compartments. The algorithm brought each scan into an approximately rigid register with the MNI template, making subsequent transformations more robust. The standard co-registration algorithm in SPM12 was then applied to co-register each CT scan with its coupled T1-weighted MRI, which enabled automatic replication of each subsequent transformation of the T1-weighted images with their corresponding CT scans. Standard segmentation and normalisation routines with default parameters were then applied to the T1-weighted images to create segmented images in native space for each of the six standard tissue classes, combined with a set of non-linear parameters, to transform the resulting segments into MNI space. These parameters were then used to transform the white matter and grey matter compartments of each T1-weighted image and the corresponding CT scan into normalised MNI space. The location of the electrodes in MNI space was determined by displaying the normalised T1-weighted and CT images together in triplanar view using the SPM12's 'check registration' module. The location of the centre of each electrode was visually judged by two independent observers to lie within the grey matter of the medial wall. By locating the electrode contacts in MNI space after normalisation, the potential bias due to anatomical differences between subjects was minimised during manual labelling.

2.5 *Local disruptive mapping*

A total of 477 stimulation locations were extracted, covering a larger region than the original study. For each location a corresponding image (1.5 × 1.5 ×1.5 mm sampled) was generated with intensity of zero at all locations except the stimulated location where the intensity was one. Each image was convolved with a 3D Gaussian 10 mm full-width-half-maximum (FWHM) kernel (truncated at 90% mass) to enable modelling of spatial uncertainty in the location of the stimulation and approximate the local distribution of focally induced disruption. The resultant image contained a single Gaussian located at the stimulation point, representing location uncertainty as the density of this spatial distribution across all voxels. This approach facilitated group analysis of sparse data accounting for between-subject variation in functional-anatomical relationships not captured by anatomical registrations, analogous to the approach used in meta-analytic modelling of functional activation data (Eickhoff et al., 2009; Eickhoff et al., 2012). The kernel size used here was guided by empirical studies on spatial uncertainty modelling of functional neuroanatomical data (Eickhoff et al., 2009). Manipulating the kernel width from 4 to 16 mm in increments of 2 mm, Trevisi et al. (2018) reported similar results, suggesting that the choice of the kernel size is not a critical step in the analysis. We do not claim that our choice of kernel width is optimal or generally prescribable, though it can be—as here—empirically informed by data from other modalities. Note also that the kernel size is dominated by plausible inter-subject variability rather than the comparatively much smaller scale of current spread (Chaturvedi et al., 2013).

Trevisi et al. (2018)'s focus on the rostro-caudal organisation of the medial wall motivated them to collapse the data across other planes. Here the enhanced efficiency of our approach allowed us to investigate bilateral effects. Data were masked by applying a threshold where electrode density was greater than 0.00001 to exclude areas with poor sampling. The subsequent mask was confined to the frontal medial wall, and extended laterally 22 mm to encompass its depths. For each behavioural condition of interest, stimulation images were entered into a voxel-wise repeated-measures general linear model with electrode density as the dependent variable and subject and the binary behavioural effect as the independent variables. Within-subject non-sphericity of errors was accounted for using standard procedures (Friston et al., 2002). A planned one-tailed voxel-wise t-test of each behavioural condition was performed and thresholded at $p < 0.05$ FWE (peak voxel) to account for multiple comparisons.

2.6 *Connective disruptive mapping*

We used large-scale high-resolution diffusion tensor imaging from the Human Connectome Project (HCP) to derive white matter connectivity matrices (Glasser et al., 2013; Sotiropoulos et al., 2013). The imaging acquisition protocols for the HCP are described elsewhere (Glasser et al., 2013). In total, data from 945 participants were deemed suitable for analysis. We used FSL pre-processed diffusion data supplied by the HCP working group (Glasser et al., 2013). In brief, this pre-processing included b0 signal intensity normalization across the six-diffusion series, and correction for echoplanar imaging distortion, eddy current and subject motion distortion, and gradient nonlinearities. Registration of the diffusion images to the native T1-weighted structural space enabled non-linear registration to $2 \times 2 \times 2$ mm sampled isotropic MNI template space via a deformation field derived with FNIRT from each individual's T1-weighted image. This included BEDPOSTX processing with the default deconvolution model using sticks with a range of diffusivities.

Probabilistic tractography was applied to the diffusion data to derive local fibre orientation information. Tractography was performed using the GPU Bayesian implementation of probtrackx2 (Hernandez-Fernandez et al., 2019), to derive a network representation of white matter structural connections across all grey matter voxels at $2 \times 2 \times 2$ mm resolution for each of the 945 participants. Our processing parameters included passing a grey matter mask, 5000 samples, a curvature threshold of 0.2, 2000 steps with a steplength of 0.5 mm, and a subsidiary fibre volume fraction threshold of 0.01, with normalisation by the participant waytotal. The waytotal is the total number of generated tracts that satisfy the inclusion/exclusion mask criteria: normalising by it scales the estimated values to the local tract density, enabling better accounting of connectivity variations with distance. The probabilistic tractography approach and implementation are described elsewhere (Behrens et al., 2003; Hernandez-Fernandez et al., 2019).

Having derived a white matter tractography network of each grey matter voxel, we averaged the streamline samples across all 945 patients to yield a large adjacency matrix, which could be incorporated into an undirected weighted graph. The graph comprised 125760 individual grey matter voxels, with 7907725920 unique edges weighted by the mean normalised-streamline value, subsequently used to infer the strength of the structural connection between grey matter voxels.

Note that optimal approach to deriving white matter, amongst other, connectivity maps is a subject of intense study (Jones, 2010; Jones & Cercignani, 2010): for our purposes a widely used exemplar is sufficient. Any alternative may be substituted, including subject-specific maps derived from individual tractographic or resting state imaging. Note also that the distance normalisation employed here deliberately magnifies the remote effects it is the objective of this approach to reveal. Others may choose to forego this step, or to explore its effects on the downstream statistics.

Given the set of stimulation coordinates, we used this structural connectome to reconstruct brain maps depicting *connection strength from a stimulation seed point to all other grey matter voxels*. The connection strength (edge weight) from the stimulated voxel to each other grey matter voxel was rendered as an image volume—one for each stimulation location—for subsequent analysis. We used intensity clamping outside 0.1 and 99.9% of the intensity cumulative density to eliminate the influence of presumably spurious extreme values.

As the resulting maps are already dense (in contrast with the sparse focal disruption maps), for the dysconnectomic analysis we applied a smaller smoothing kernel of 6 mm FWHM prior to assessing association for each behavioural outcome with the same statistical design as used above for local disruption mapping. Again, a voxel-wise repeated-measures general linear model with subject and the binary behavioural effect as factors was used, and non-sphericity of errors was accounted for (Friston et al., 2002). Planned one-tailed voxel-wise t-tests were performed and thresholded at $p < 0.05$ FWE (peak voxel) – now revealing dysconnectome maps.

2.8 *Visualisation*

Visualisation was done using the SurfIce toolbox (https://www.nitrc.org/projects/surfice). We used FSL's HCP1065 standard-space FA atlas to generate a background FA template (Yeh et al., 2018) converted into a mesh using SurfIce's volume-to-mesh function.

The local disruption and normalised disconnectome maps were visualised at the PWE corrected threshold in the mid-sagittal, lateral, superior, and inferior views. The mid-sagittal laterality was determined by the x coordinated value of global maxima of each map, so that for x < 0 the left hemisphere is shown, whereas for x ≥0 the right hemisphere is shown. The nearest grey matter location to the maxima of each statistical cluster was determined using SPM12's neuromorphometrics grey matter atlas for cortical regions. For cerebellar regions, we used a well-established functional parcellation into distinct networks based on resting state connectivity (Marek et al. (2018)). To illustrate the extent of the effects, the local disruption and normalised dysconnectome maps were also visualised at a lower statistical threshold ($p = 0.001$ uncorrected, t = 3.11) and are shown in Supplementary Figures S1 and S2 respectively.

## 3. Results

3.1 *Local disruptive mapping*

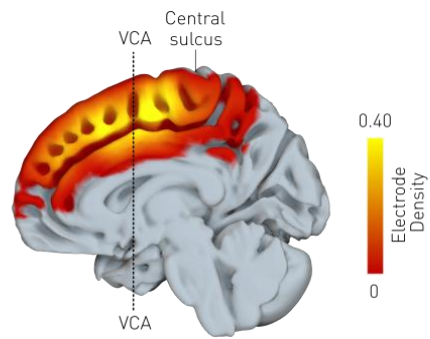

**Figure 2. Distribution of stimulation locations over the frontal medial wall.** The mean stimulation density (smoothed by a 10mm FWHM Gaussian kernel) is shown overlaid onto the FA template. The colourbar shows the density of electrode locations.

A total of 477 disruption sites confined to the medial frontal wall were obtained from 37 patients, providing good sampling coverage across the medial wall (Figure 2, Supplementary Table T2).

Positive motor responses—the most common stimulation-induced behaviour—were observed in 153 stimulations (32%) (Figure 3, Supplementary Table T3). They were associated with disruptions in the right SMA and the precentral gyrus bilaterally. Negative motor responses were observed in 41 stimulations (8%). They were associated with disruptions in the SMA and the right middle segment of the superior frontal gyrus. Sensory responses were observed in 46 stimulations (9%). They were associated with disruptions in the left middle cingulate gyrus, and the middle segment of the precentral gyrus (bilaterally). Speech disturbances were observed in 46 stimulations (9%). They were associated with disruptions of the left superior frontal gyrus and pre-SMA. No-responses (silent disruptions) were observed in 243 stimulations (51%). They were associated with disruptions of the middle cingulate gyrus and the middle segment of the superior frontal gyrus. Results at the uncorrected threshold, are available in Supplementary Figure F1.

To verify the focal results are not dependent on the predetermined smoothing kernel, we performed a sensitivity analysis using 8, 10, and 12 kernels (Supplementary Figure F3), revealing no significant differences in the location and spread of the effects.

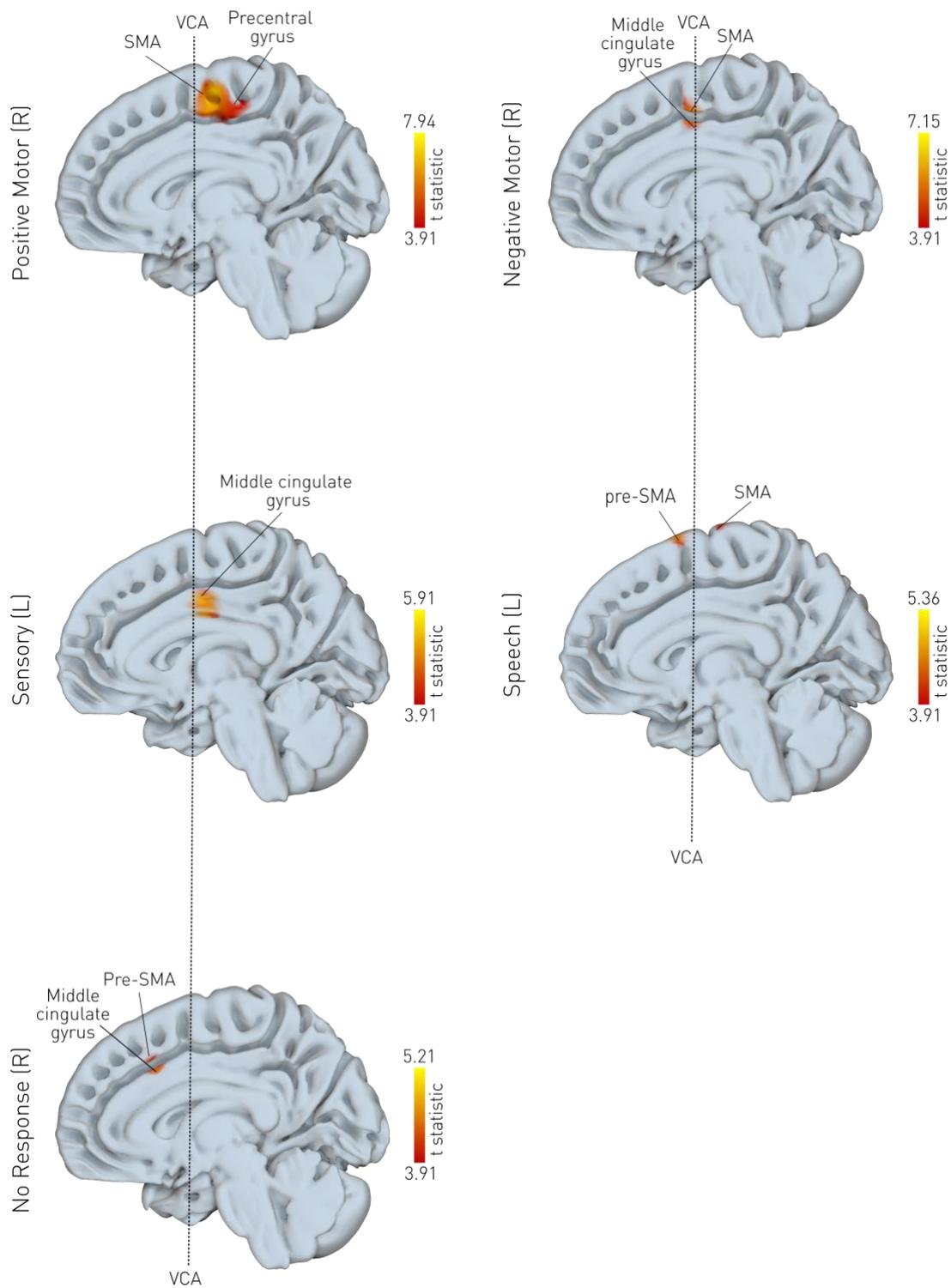

**Figure 3. Local disruptive mapping of behaviour**. For each MNI voxel, a planned t-contrast was performed. Only voxels surviving the $p < 0.05$ FWE-corrected threshold are shown, overlaid on the mid-sagittal plane, where higher t-statistics (brighter colour) represent a stronger association between the electrode density value and the observed behaviour. R = right; L = left; (pre-)SMA = supplementary motor area.

## 3.2 *Dysconnectomic disruptive mapping*

Positive motor responses were localized to 14 clusters (Figure 4, supplementary Table T4). Cortical grey matter regions included the precentral gyrus, superior parietal lobule, superior frontal gyrus, posterior insula, central operculum, and the superior occipital gyrus. Deep grey matter regions included the thalamus, ventral diencephalon, caudate nucleus, and pallidum. Cerebellar connectivity was evident in regions falling within the fronto-parietal, foot, and hand sensorimotor cerebellar networks (Marek et al., 2018). At the uncorrected threshold, connectivity was also evident in the mid-brain, pons, and medulla, as can be appreciated in Supplementary Figure F2. Negative motor responses were localized to 12 clusters, with the global maximum located in the superior parietal lobule. Cortical grey matter regions included the superior parietal lobule, SMA, superior and medial frontal gyrus, and precentral gyrus. Significant clusters were also found in the precuneus, and the cerebellar fronto-parietal and dorsal attention networks. Sensory responses were localized to 5 clusters, with the global maximum falling within the precentral gyrus. Significant areas were also obtained in the middle cingulate cortex, thalamus, the fronto-parietal cerebellar network, and the medulla. The absence of a response to disruption was associated with 7 clusters, with global maxima falling in the medial segment of the superior frontal gyrus. Other areas included the middle frontal gyrus, medial orbital gyrus, precuneus, angular gyrus and the dorsal attention cerebellar network.

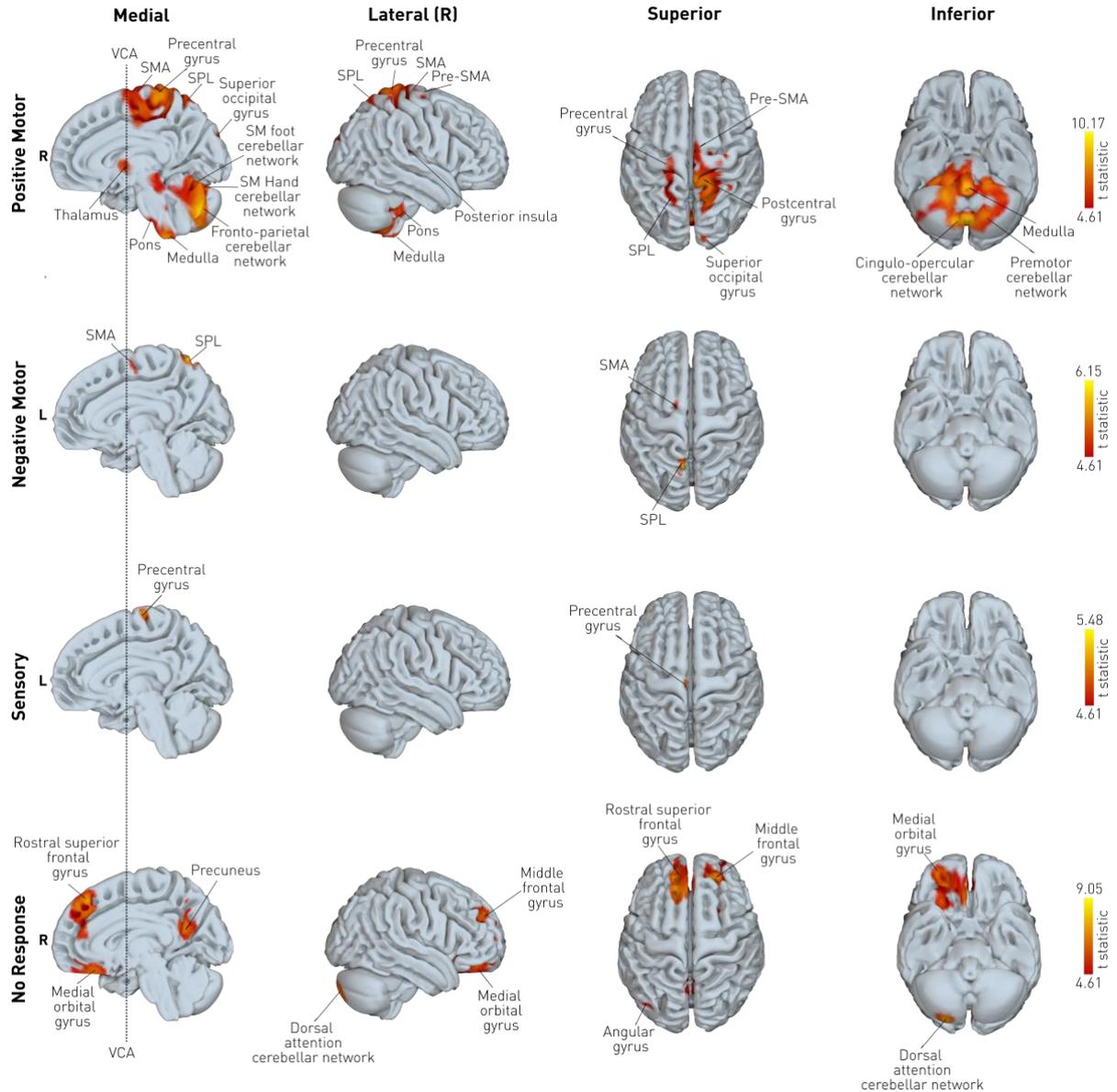

**Figure 4. Connective disruptive maps of behaviour.** For each MNI voxel, a planned t-contrast was performed. Only voxels surviving the $p < 0.05$ FWE-corrected threshold are shown, overlaid on the mid-sagittal, lateral, superior, and inferior planes, where higher t-statistics (brighter colour) represent a stronger association between the connectivity value and the observed behaviour. Cerebellar subregions are labelled with reference to a priori known cortical network associations (Marek et al., 2018). R = right; L = left; SMA = supplementary motor area; SPL = superior parietal lobule, SM = sensorimotor).

## 4. Discussion

We have presented a framework for local and connective spatial inference with sparsely sampled focal disruptive data and applied it to transient direct cortical electrical stimulation of the medial frontal wall. Here we review the characteristics of our approach, and examine the empirical results drawn from its application to illuminating the local and distributed organisation of the medial frontal wall.

4.1 *Spatial inference from sparsely sampled discrete data*

Inferring a dense map from sparsely sampled disruptive data inevitably implies interpolation between unsampled regions of the target space guided by the chosen method's inductive bias. ROI-based analysis assumes all voxels within any given region are equivalent, changing function with implausible abruptness across regions (Glasser et al., 2016). It yields maps that are discretised not by the data themselves, but on an assumption the inference rests on. If there is a spatial scale at which this assumption is safe, it is yet to be discovered: current parcellations of the brain are based on arbitrarily discretised continuities. By contrast, the alternative approach we pursue here merely assumes that the underlying substrate can be modelled by a random Gaussian field. The one heuristic—the width of the Gaussian with which point data is convolved—can be both empirically informed by data from other modalities and corresponds to the readily intelligible notion of chosen scale of spatial analysis.

Representing the data in Gaussian-convolved form furthermore enables the application of voxel-wise mass-univariate methods whose sensitivity and statistical efficiency makes them preferable where, as here, a given locus may licitly be evaluated independently. SPM's established statistical framework combines flexibility in the framing of voxel-wise statistical hypotheses with great fidelity in the context of marked noise. Where multiple loci are disrupted together, as in natural lesions, the assumptions of benignly structured local dependence are broken, and mass univariate inference is not valid, whether with dense or sparse data, and a multivariate approach becomes necessary (Mah et al., 2014; Xu et al., 2018).

Representing stimulation data as a disrupted cortical network of connections—a dysconnectome—naturally extends the evaluated neural support across the brain. Connectivity need neither be structural—functional or meta-analytic indices, for example, are equally applicable—nor unitary: multiple dysconnectome maps could theoretically be compared, within the same or parallel models. Metrics of connectivity may be transformed to focus attention on different spatial scales of interaction, such as the distance-based normalisation employed in our study to highlight long-range connectivity. Though group inference requires a common space (at least implicitly), a dysconnectome may be estimated from individual rather than template connectivity data. Dysconnectome maps are open to mass-univariate inference, with the caveat that the interactions between distributed areas are left unmodelled, just as they are when multiple regions are identified in functional imaging. A full network analysis requires an explicit model of the interactions between regions, for which a dysconnectome map may serve as an initial feature selection step.

Our approach is analogous to lesion disconnectome mapping, where a lesion is projected along the white matter tracts it encloses, yielding a probabilistic representation of disrupted white matter pathways delineated up to the boundaries of the cortical areas they connect (Foulon et al., 2018). The difference is that our lesions are single point loci falling entirely within grey matter, and the projection we seek is to connected grey matter areas, not to the underlying white matter. The inferred topology is then primarily in terms of grey rather than white matter anatomy, defined by any chosen mechanism of connectivity.

An obvious question is the optimal stage at which connective inference should be performed. One could estimate a dysconnectome from the regions a prior local analysis identifies as significantly associated with the behaviour at the group level. This alternative is suboptimal for two reasons. First, it precludes modelling of individual variations in the distributed substrate that individual connectivity maps could provide. More importantly, it diminishes the spatial resolution of the inference: two sites of disruption too close to be resolved within a focal map may exhibit connectivity profiles that successfully distinguish them at the dysconnectome level. One may conceive of a dysconnectome as a projection into a higher dimensional anatomical space, akin to an anatomical support vector machine, that enhances the separability of the underlying patterns of dependence.

These generalities apply to the particulars of direct cortical electrical stimulation data. Note that inter-subject variability, even after precise non-linear registration, likely dwarfs the effects of local current spread, rendering physics-informed spatial priors unnecessary (Chaturvedi et al., 2013). The method of inference naturally cannot overcome the limitations of the data themselves, which here include sparse and clinically biased sampling, non-linearities of induced effects, clinically-determined variations in stimulation parameters, interactions – both enduring (e.g., coexisting structural lesions) and dynamic (e.g., after-discharges) – with the pathology that motivates intracranial study, confounding from other interventions (e.g., anti-epileptic drugs), and the relatively narrow repertoire of behaviours the clinical setting permits us to evaluate (Theodore, 1988). That the disruption is localised to a comparatively small volume, however, is a crucial advantage over natural lesions, where the distributed nature of the damage enormously complicates spatial inference, and where many of the foregoing limitations apply equally.

Clinical constraints make it inevitable that only a comparatively small number of areas can be sampled in any one patient. Just as with structural lesions, such as ischaemic stroke, where each lesion is associated with a single patient and is treated as an independent datum, so here each disruption site is treated as a single lesion and as an independent datum. Where multiple sites are drawn from the same patient, however, each observation is no longer independent, which needs to be accounted for in the statistical model. Here we employ a single-level repeated measures design, where linear variations in the mean of individual patients is absorbed by a subject-level regressor, and the non-sphericity of the error induced by within-subject correlations is modelled by whitening the design matrix (Friston et al., 2002). This manoeuvre, standard in SPM, does not change the degrees of freedom of the model, unlike alternatives such as the approximate Greenhouse-Geisser correction (Greenhouse & Geisser, 1959). A valid alternative is to adopt an approximate hierarchical, two-level design where contrasts from individual subject-specific models at the first level are evaluated by a one-sample t-test at a second level. We favour the former approach owing to its flexibility and power in the context of the imbalanced data the present clinical context tends to compel.

Special attention must be given to the heterogeneity of the neural effects of electrical stimulation. The effects at the single neuron level may be excitatory or inhibitory, varying both over time and distance, and need not correspond to effects at larger scales that inevitably depend both on individual neuronal responses and the configuration—local and remote—of physiological excitatory and inhibitory neuronal functions. Nor can a predominance of induced functional excitation or inhibition be inferred from the elicited behaviour: the interruption of movement need not be neurally inhibitory even if it is behaviourally so, and vice versa. Equally, a movement may be disrupted not because it has lost its substrate but because a competing substrate has been driven to supplant it; and may fail to manifest positively not because the substrate is not excited but because the complexity of the movement demands a pattern of excitation too elaborate for electrical stimulation to induce. These considerations are especially pertinent to the interpretation of rostral effects where negative effects or silence reign: we cannot conclude that these regions are inhibitory, only that the behaviour that depends on them may be competitive or complex. In short, electrical stimulation is correctly viewed as transient, heterogeneous disruption, of value in localisation but not necessarily the more detailed characterisation of the neural substrate.

4.2 *Focal mapping of the medial frontal wall*

Our analysis broadly replicates the maps obtained from the original permutation-based voxel-wise analysis of the same data (Trevisi et al., 2018). A rostro-caudal organisation of behavioural complexity is observed (Badre & Nee, 2018; Thiebaut de Schotten et al., 2017; Badre, 2008; Nachev et al. 2008; Koechlin et al., 2003; Picard & Strick, 1996), with positive motor responses caudal to negative motor, speech, idiosyncratic phenomena ('other' responses), and absence of any response. In presupposing a task that is contextually interrupted, negative responses can reasonably be expected to be more complex in their condition-action association than positive ones. Speech elicited a locally distributed pattern likely reflecting the complex compositionality of the task, including relatively low-level aspects of articulation (Chee et al., 1997; Chang et al. 2020). The rarity of idiosyncratic, 'other' responses makes the interpretation of the highly localised effect we observe here difficult: larger samples of these rostral areas, in combination with more elaborate tasks are required to cast reliable light on them.

The crudity of the behavioural labels here is a reminder anatomical inference is not to functions but to behaviours speculated to depend on them. Obviously, no function corresponds to the "silence" observed across of the vast expanse of the medial wall where no response of any kind could be elicited. The inference to be drawn is not that this region does nothing but that the behavioural tasks commonly deployed to explore its function in the clinical setting lack range and specificity.

4.3 *Connectomic mapping of the medial frontal cortex with transient dysconnectomes*

Projecting disrupted cortical areas to their grey-matter connections reveals a much richer picture of the underlying neural substrate. Positive motor responses are centred on the supplementary motor area, with extensive precentral gyrus involvement plausibly reflecting strong connectivity with primary motor cortex. There are extensive projections to subcortical targets including thalamus, basal ganglia, pons, medulla, and cerebellum. Negative motor responses show more rostral weighting on the medial wall, and weaker connections to precentral and deep areas. By contrast, SPL involvement was more prominent here, giving context to recent evidence for direct parietal-motor functional connectivity in the region (Cattaneo et al. 2020). Sensory responses highlight the middle cingulate gyrus in keeping with its known behavioural associations (e.g., Hadland et al., 2003; Vogt, 2016; Lim et al., 1994, 1996), and connectivity patterns (Beckmann et al. 2009). The region we identify is plausibly part of the caudal cingulate premotor area involved in the multisensory orientation of the head and body in space (Vogt, 2016).

Special note should be made of cerebellar connectivity, structured by its recently described network organisation (Marek et al., 2018). Here positive motor responses engage multiple areas including cerebellar nodes of the fronto-parietal, hand and foot sensorimotor, and dorsal and ventral attention networks. Negative responses are by contrast dominated by the fronto-parietal network, as befits their more complex generative context. Sensory responses overlap with a component of the cingulo-opercular network, in line with the strong association with the mid-cingulate. The association of silent responses with cerebellar connectivity within the dorsal attention network, shown to play a role in memory and attention tasks previously thought to be dominated by cortical regions (Brissenden et al., 2016), suggests the category of tasks more likely to be eloquently modulated by disruption of the rostral medial wall. Our findings highlight the strongly structured spatial organisation of the cerebellum, inviting future research to further delineate its interaction with cortical and subcortical regions.

4.4 *Conclusion*

We present—and apply to the medial frontal wall—a new random field-based approach to drawing spatial inferences about the focal and distributed functional organisation of the brain from sparsely sampled disruptive data. Our approach combines minimal anatomical and physiological assumptions with an efficient, principled framework for establishing disruption-behavioural associations. Applied to the medial wall, it reveals marked differences between focal and distributed maps, even in the context of relatively constrained spatial sampling, with implications for our understanding of the functional organisation of the region, and—more generally—the optimal path to integrating local and distributed information in our models of the brain.

**References**

Adolphs, R. (2016). Human Lesion Studies in the 21st Century. *Neuron*, *90*(6), 1151–1153. https://doi.org/10.1016/j.neuron.2016.05.014

Alivisatos, A. P., Chun, M., Church, G. M., Greenspan, R. J., Roukes, M. L., & Yuste, R. (2012). The brain activity map project and the challenge of functional connectomics. *Neuron*, *74*(6), 970–974. https://doi.org/10.1016/j.neuron.2012.06.006

Amunts, K., Schleicher, A., Bürgel, U., Mohlberg, H., Uylings, H. B., & Zilles, K. (1999). Broca's region revisited: Cytoarchitecture and intersubject variability. The Journal of Comparative Neurology, *412*(2), 319–341. https://doi.org/10.1002/(sici)1096-9861(19990920)412:2<319::aid-cne10>3.0.co;2-7

Badre, D. (2008). Cognitive control, hierarchy, and the rostro-caudal organization of the frontal lobes. Trends in Cognitive Sciences, *12*(5), 193–200. https://doi.org/10.1016/j.tics.2008.02.004

Badre, D., & Nee, D. E. (2018). Frontal Cortex and the Hierarchical Control of Behavior. Trends in Cognitive Sciences, *22*(2), 170–188. https://doi.org/10.1016/j.tics.2017.11.005

Beckmann, M., Johansen-Berg, H., & Rushworth, M. F. S. (2009). Connectivity-Based Parcellation of Human Cingulate Cortex and Its Relation to Functional Specialization. *Journal of Neuroscience*, *29*(4), 1175–1190. https://doi.org/10.1523/JNEUROSCI.3328-08.2009

Behrens, T. E. J., Woolrich, M. W., Jenkinson, M., Johansen-Berg, H., Nunes, R. G., Clare, S., Matthews, P. M., Brady, J. M., & Smith, S. M. (2003). Characterization and propagation of uncertainty in diffusion-weighted MR imaging. *Magnetic Resonance in Medicine*, *50*(5), 1077–1088. https://doi.org/10.1002/mrm.10609

Boes, A. D., Prasad, S., Liu, H., Liu, Q., Pascual-Leone, A., Caviness, V. S., & Fox, M. D. (2015). Network localization of neurological symptoms from focal brain lesions. Brain, *138*(10), 3061–3075. https://doi.org/10.1093/brain/awv228

Brissenden, J. A., Levin, E. J., Osher, D. E., Halko, M. A., & Somers, D. C. (2016). Functional Evidence for a Cerebellar Node of the Dorsal Attention Network. The Journal of Neuroscience, *36*(22), 6083–6096. https://doi.org/10.1523/JNEUROSCI.0344-16.2016

Catani, M., Dell'Acqua, F., Bizzi, A., Forkel, S. J., Williams, S. C., Simmons, A., Murphy, D. G., & Thiebaut de Schotten, M. (2012). Beyond cortical localization in clinico-anatomical correlation. Cortex, *48*(10), 1262–1287. https://doi.org/10.1016/j.cortex.2012.07.001

Cattaneo, L., Giampiccolo, D., Meneghelli, P., Tramontano, V., & Sala, F. (2020). Cortico-cortical connectivity between the superior and inferior parietal lobules and the motor cortex assessed by intraoperative dual cortical stimulation. *Brain Stimulation*, *13*(3), 819–831. https://doi.org/10.1016/j.brs.2020.02.023

Chang, E. F., Kurteff, G., Andrews, J. P., Briggs, R. G., Conner, A. K., Battiste, J. D., & Sughrue, M. E. (2020). Pure Apraxia of Speech After Resection Based in the Posterior Middle Frontal Gyrus. *Neurosurgery*, *87*(3), E383–E389. https://doi.org/10.1093/neuros/nyaa002

Chaturvedi, A., Luján, J. L., & McIntyre, C. C. (2013). Artificial neural network based characterization of the volume of tissue activated during deep brain stimulation. Journal of Neural Engineering, *10*(5), 056023. https://doi.org/10.1088/1741-2560/10/5/056023

Chee, M. W., So, N. K., & Dinner, D. S. (1997). Speech and the dominant superior frontal gyrus: Correlation of ictal symptoms, EEG, and results of surgical resection. *Journal of Clinical Neurophysiology: Official Publication of the American Electroencephalographic Society*, *14*(3), 226–229. https://doi.org/10.1097/00004691-199705000-00007

Corrivetti, F., de Schotten, M. T., Poisson, I., Froelich, S., Descoteaux, M., Rheault, F., & Mandonnet, E. (2019). Dissociating motor-speech from lexico-semantic systems in the left frontal lobe: Insight from a series of 17 awake intraoperative mappings in glioma patients. Brain Structure & Function, *224*(3), 1151–1165. https://doi.org/10.1007/s00429-019-01827-7

DeMarco, A. T., & Turkeltaub, P. E. (2020). Functional anomaly mapping reveals local and distant dysfunction caused by brain lesions. *NeuroImage*, *215*, 116806. https://doi.org/10.1016/j.neuroimage.2020.116806

Desmurget, M., & Sirigu, A. (2015). Revealing humans' sensorimotor functions with electrical cortical stimulation. Philosophical Transactions of the Royal Society B: Biological Sciences, *370*(1677), 20140207. https://doi.org/10.1098/rstb.2014.0207

Eickhoff, S. B., Bzdok, D., Laird, A. R., Kurth, F., & Fox, P. T. (2012). Activation likelihood estimation meta-analysis revisited. *NeuroImage*, *59*(3), 2349–2361. https://doi.org/10.1016/j.neuroimage.2011.09.017

Eickhoff, S. B., Laird, A. R., Grefkes, C., Wang, L. E., Zilles, K., & Fox, P. T. (2009). Coordinate-based activation likelihood estimation meta-analysis of neuroimaging data: A random-effects approach based on empirical estimates of spatial uncertainty. *Human Brain Mapping*, *30*(9), 2907–2926. https://doi.org/10.1002/hbm.20718

Filevich, E., Kühn, S., & Haggard, P. (2012). Negative motor phenomena in cortical stimulation: Implications for inhibitory control of human action. *Cortex; a Journal Devoted to the Study of the Nervous System and Behavior*, *48*(10), 1251–1261. https://doi.org/10.1016/j.cortex.2012.04.014

Fischl, B., van der Kouwe, A., Destrieux, C., Halgren, E., Ségonne, F., Salat, D. H., Busa, E., Seidman, L. J., Goldstein, J., Kennedy, D., Caviness, V., Makris, N., Rosen, B., & Dale, A. M. (2004). Automatically parcellating the human cerebral cortex. *Cerebral Cortex (New York, N.Y.: 1991)*, *14*(1), 11–22. https://doi.org/10.1093/cercor/bhg087

Foulon, C., Cerliani, L., Kinkingnéhun, S., Levy, R., Rosso, C., Urbanski, M., Volle, E., & Thiebaut de Schotten, M. (2018). Advanced lesion symptom mapping analyses and implementation as BCBtoolkit. GigaScience, 7(3), giy004. https://doi.org/10.1093/gigascience/giy004

Friston, K. J., Holmes, A. P., Worsley, K. J., Poline, J.-P., Frith, C. D., & Frackowiak, R. S. J. (1994). Statistical parametric maps in functional imaging: A general linear approach. *Human Brain Mapping*, *2*(4), 189–210. https://doi.org/10.1002/hbm.460020402

Friston, K. J., Glaser, D. E., Henson, R. N. A., Kiebel, S., Phillips, C., & Ashburner, J. (2002). Classical and Bayesian inference in neuroimaging: Applications. NeuroImage, *16(2)*, 484–512. https://doi.org/10.1006/nimg.2002.1091

Glasser, M. F., Coalson, T. S., Robinson, E. C., Hacker, C. D., Harwell, J., Yacoub, E., Ugurbil, K., Andersson, J., Beckmann, C. F., Jenkinson, M., Smith, S. M., & Van Essen, D. C. (2016). A multi-modal parcellation of human cerebral cortex. *Nature*, *536*(7615), 171–178. https://doi.org/10.1038/nature18933

Glasser, M. F., Sotiropoulos, S. N., Wilson, J. A., Coalson, T. S., Fischl, B., Andersson, J. L., Xu, J., Jbabdi, S., Webster, M., Polimeni, J. R., Van Essen, D. C., Jenkinson, M., & WU-Minn HCP Consortium. (2013). The minimal preprocessing pipelines for the Human Connectome Project. *NeuroImage*, *80*, 105–124. https://doi.org/10.1016/j.neuroimage.2013.04.127

Goldstein, J. M., Goodman, J. M., Seidman, L. J., Kennedy, D. N., Makris, N., Lee, H., Tourville, J., Caviness, V. S., Faraone, S. V., & Tsuang, M. T. (1999). Cortical abnormalities in schizophrenia identified by structural magnetic resonance imaging. *Archives of General Psychiatry*, *56*(6), 537–547. https://doi.org/10.1001/archpsyc.56.6.537

Greenhouse, S. W., & Geisser, S. (1959). On methods in the analysis of profile data. *Psychometrika*, *24*(2), 95–112. https://doi.org/10.1007/BF02289823

Hadland, K. A., Rushworth, M. F. S., Gaffan, D., & Passingham, R. E. (2003). The effect of cingulate lesions on social behaviour and emotion. *Neuropsychologia*, *41*(8), 919–931. https://doi.org/10.1016/s0028-3932(02)00325-1

Hernandez-Fernandez, M., Reguly, I., Jbabdi, S., Giles, M., Smith, S., & Sotiropoulos, S. N. (2019). Using GPUs to accelerate computational diffusion MRI: From microstructure estimation to tractography and connectomes. *NeuroImage*, *188*, 598–615. https://doi.org/10.1016/j.neuroimage.2018.12.015

Jha, A., Diehl, B., Scott, C., McEvoy, A. W., & Nachev, P. (2016). Reversed Procrastination by Focal Disruption of Medial Frontal Cortex. *Current Biology*, *26*(21), 2893–2898. https://doi.org/10.1016/j.cub.2016.08.016

Jha, A., Teotonio, R., Smith, A.-L., Bomanji, J., Dickson, J., Diehl, B., Duncan, J. S., & Nachev, P. (2020). Metabolic lesion-deficit mapping of human cognition. Brain, *143*(3), 877–890. https://doi.org/10.1093/brain/awaa032

Jones, D. K. (2010). Challenges and limitations of quantifying brain connectivity in vivo with diffusion MRI. Imaging in Medicine, *2*(3), 341.

Jones, D. K., & Cercignani, M. (2010). Twenty-five pitfalls in the analysis of diffusion MRI data. NMR in Biomedicine, *23*(7), 803–820. https://doi.org/10.1002/nbm.1543

Kim, J.-H., Lee, J.-M., Jo, H. J., Kim, S. H., Lee, J. H., Kim, S. T., Seo, S. W., Cox, R. W., Na, D. L., Kim, S. I., & Saad, Z. S. (2010). Defining functional SMA and pre-SMA subregions in human MFC using resting state fMRI: Functional connectivity-based parcellation method. *NeuroImage*, *49*(3), 2375–2386. https://doi.org/10.1016/j.neuroimage.2009.10.016

Koechlin, E., Ody, C., & Kouneiher, F. (2003). The architecture of cognitive control in the human prefrontal cortex. Science (New York, N.Y.), *302*(5648), 1181–1185. https://doi.org/10.1126/science.1088545

Kovac, S., Scott, C. A., Maglajlija, V., Toms, N., Rodionov, R., Miserocchi, A., McEvoy, A. W., & Diehl, B. (2014). Comparison of bipolar versus monopolar extraoperative electrical cortical stimulation mapping in patients with focal epilepsy. *Clinical Neurophysiology*, *125*(4), 667–674. https://doi.org/10.1016/j.clinph.2013.09.026

Lim, S. H., Dinner, D. S., & Lüders, H. O. (1996). Cortical stimulation of the supplementary sensorimotor area. Advances in Neurology, 70, 187–197.

Lim, S. H., Dinner, D. S., Pillay, P. K., Lüders, H., Morris, H. H., Klem, G., Wyllie, E., & Awad, I. A. (1994). Functional anatomy of the human supplementary sensorimotor area: Results of extraoperative electrical stimulation. Electroencephalography and Clinical Neurophysiology, 91(3), 179–193. https://doi.org/10.1016/0013-4694(94)90068-X

Lüders, H. O., Dinner, D. S., Morris, H. H., Wyllie, E., & Comair, Y. G. (1995). Cortical electrical stimulation in humans. The negative motor areas. *Advances in Neurology*, *67*, 115–129.

Mah, Y.-H., Husain, M., Rees, G., & Nachev, P. (2014). Human brain lesion-deficit inference remapped. *Brain: A Journal of Neurology*, *137*(Pt 9), 2522–2531. https://doi.org/10.1093/brain/awu164

Marek, S., Siegel, J. S., Gordon, E. M., Raut, R. V., Gratton, C., Newbold, D. J., Ortega, M., Laumann, T. O., Adeyoma, B., Miller, D. B., Zheng, A., Lopez, K. C., Berg, J. J., Coalson, R. S., Nguyen, A. L., Dierker, D., Van, A. N., Hoyt, C. R., McDermott, K. B., … Dosenbach, N. U. F. (2018). Spatial and Temporal Organization of the Individual Human Cerebellum. *Neuron*, *100*(4), 977-993.e7. https://doi.org/10.1016/j.neuron.2018.10.010

Nachev, P., Kennard, C., & Husain, M. (2008). Functional role of the supplementary and pre-supplementary motor areas. *Nature Reviews Neuroscience*, *9*(11), 856–869. https://doi.org/10.1038/nrn2478

Nowell, M., Rodionov, R., Diehl, B., Wehner, T., Zombori, G., Kinghorn, J., Ourselin, S., Duncan, J., Miserocchi, A., & McEvoy, A. (2014). A Novel Method for Implementation of Frameless StereoEEG in Epilepsy Surgery. *Operative Neurosurgery*, *10*(4), 525–534. https://doi.org/10.1227/NEU.0000000000000544

Nudo, R. J., & Friel, K. M. (1999). Cortical plasticity after stroke: Implications for rehabilitation. *Revue Neurologique*, *155*(9), 713–717.

Penfield, W. (1950). The supplementary motor area in the cerebral cortex of man. *Archiv Fur Psychiatrie Und Nervenkrankheiten, Vereinigt Mit Zeitschrift Fur Die Gesamte Neurologie Und Psychiatrie*, *185*(6–7), 670–674. https://doi.org/10.1007/BF00935517

Picard, N., & Strick, P. L. (1996). Motor areas of the medial wall: A review of their location and functional activation. *Cerebral Cortex (New York, N.Y.: 1991)*, *6*(3), 342–353. https://doi.org/10.1093/cercor/6.3.342

Rorden, C., & Karnath, H.-O. (2004). Using human brain lesions to infer function: A relic from a past era in the fMRI age? Nature Reviews. Neuroscience, *5*(10), 813–819. https://doi.org/10.1038/nrn1521

Roux, F.-E., Boulanouar, K., Lotterie, J.-A., Mejdoubi, M., LeSage, J. P., & Berry, I. (2003). Language functional magnetic resonance imaging in preoperative assessment of language areas: Correlation with direct cortical stimulation. Neurosurgery, *52*(6), 1335–1345; discussion 1345-1347. https://doi.org/10.1227/01.neu.0000064803.05077.40

Salvalaggio, A., De Filippo De Grazia, M., Zorzi, M., Thiebaut de Schotten, M., & Corbetta, M. (2020). Post-stroke deficit prediction from lesion and indirect structural and functional disconnection. Brain, *143*(7), 2173–2188. https://doi.org/10.1093/brain/awaa156

Sarubbo, S., Tate, M., De Benedictis, A., Merler, S., Moritz-Gasser, S., Herbet, G., & Duffau, H. (2020). Mapping critical cortical hubs and white matter pathways by direct electrical stimulation: An original functional atlas of the human brain. *NeuroImage*, *205*, 116237. https://doi.org/10.1016/j.neuroimage.2019.116237

Sarubbo, S., De Benedictis, A., Merler, S., Mandonnet, E., Balbi, S., Granieri, E., & Duffau, H. (2015). Towards a functional atlas of human white matter. Human Brain Mapping, *36*(8), 3117–3136. https://doi.org/10.1002/hbm.22832

Sotiropoulos, S. N., Jbabdi, S., Xu, J., Andersson, J. L., Moeller, S., Auerbach, E. J., Glasser, M. F., Hernandez, M., Sapiro, G., Jenkinson, M., Feinberg, D. A., Yacoub, E., Lenglet, C., Ven Essen, D. C., Ugurbil, K., & Behrens, T. E. (2013). Advances in diffusion MRI acquisition and processing in the Human Connectome Project. *NeuroImage*, *80*, 125–143. https://doi.org/10.1016/j.neuroimage.2013.05.057

Theodore, W. H. (1988). Antiepileptic drugs and cerebral glucose metabolism. Epilepsia, *29* Suppl 2, S48-55. https://doi.org/10.1111/j.1528-1157.1988.tb05797.x

Thiebaut de Schotten, M., Urbanski, M., Batrancourt, B., Levy, R., Dubois, B., Cerliani, L., & Volle, E. (2017). Rostro-caudal Architecture of the Frontal Lobes in Humans. Cerebral Cortex (New York, N.Y.: 1991), *27*(8), 4033–4047. https://doi.org/10.1093/cercor/bhw215

Trevisi, G., Eickhoff, S. B., Chowdhury, F., Jha, A., Rodionov, R., Nowell, M., Miserocchi, A., McEvoy, A. W., Nachev, P., & Diehl, B. (2018). Probabilistic electrical stimulation mapping of human medial frontal cortex. *Cortex*, *109*, 336–346. https://doi.org/10.1016/j.cortex.2018.06.015

Vogt, B. A. (2016). Midcingulate cortex: Structure, connections, homologies, functions and diseases. *Journal of Chemical Neuroanatomy*, *74*, 28–46. https://doi.org/10.1016/j.jchemneu.2016.01.010

Xu, T., Jha, A., & Nachev, P. (2018). The dimensionalities of lesion-deficit mapping. Neuropsychologia, *115*, 134–141. https://doi.org/10.1016/j.neuropsychologia.2017.09.007

Yeh, F.-C., Panesar, S., Fernandes, D., Meola, A., Yoshino, M., Fernandez-Miranda, J. C., Vettel, J. M., & Verstynen, T. (2018). Population-averaged atlas of the macroscale human structural connectome and its network topology. *NeuroImage*, *178*, 57–68. https://doi.org/10.1016/j.neuroimage.2018.05.027

Zilles, K., Schlaug, G., Matelli, M., Luppino, G., Schleicher, A., Qü, M., Dabringhaus, A., Seitz, R., & Roland, P. E. (1995). Mapping of human and macaque sensorimotor areas by integrating architectonic, transmitter receptor, MRI and PET data. *Journal of Anatomy*, *187 (Pt 3)*, 515–537.

**Statements & Declarations**

*Funding*

Michael S. Elmalem is funded by the Max Planck Institute for Human Cognitive and Brain Sciences, Leipzig, Germany. James K. Ruffle is funded by the Guarantors of Brain. Parashkev Nachev and Ashwani Jha are both funded by the Wellcome Trust (213038) and the NHR UCL Biomedical Research Centre.

*Competing Interests*

Ashwani Jha has received fees from Sanofi/Genzyme, Britannia and Novartis. Michael S. Elmalem, Hanna Moody, James K. Ruffle, Michel Thiebaut de Schotten, Patrick Haggard, Beate Diehl & Parashkev Nachev declare no competing interests.

*Author Contributions*

(1) Research Project: A. Conception, B. Data collection, C. Organisation, D. Execution; (2) Statistical analysis: A. Design, B. Execution, C. Review & critique; (3) Manuscript preparation: A. Writing of the first draft, B. Review & critique.

| | |
|---|---|
| Michael S. Elmalem: | (1)C, (1)D, (2)A, (2)B, (2)C, (3)A, (3)B |
| Hanna Moody: | (1)C, (2)A, (2)B, (3)B |
| James K. Ruffle: | (2)A, (2)B, (2)C, (3)B |
| Michel Thiebaut de Schotten: | (1)A, (3)B |
| Patrick Haggard: | (1)C, (3)B |
| Beate Diehl: | (1)A, (1)B, (1)C, (1)D, (2)C, (3)B |
| Parashkev Nachev: | (1)A, (1)B, (1)C, (2)A, (2)C, (3)A, (3)B |
| Ashwani Jha: | (1)A, (1)B, (1)C, (2)A, (2)C, (3)A, (3)B |

*Data Availability*

Coordinate and behavioural data are available from the corresponding authors on request by email. The raw imaging and video data is derived from clinical studies not licensed for public dissemination by the host institution.

*Ethics approval*

This study uses irrevocably anonymized previously published data derived from a hospital-approved service evaluation project that does not require ethical approval.

*Consent to participate*

Not applicable in the setting of irrevocably anonymized routinely collected data analysed for the primary purpose of service evaluation.

## Supplementary Table T1.

| Patient | Type | Sex | Handed | Age at onset (years) | Age at ICR (years) | Epilepsy Duration (years) | Side | Language dominance | Imaging abnormality | Ictal onset zone | Pathology (if operated) | ILAE outcome (months) |
|---|---|---|---|---|---|---|---|---|---|---|---|---|
| 1 | Grid | F | Left | 3 | 27 | 24 | Left | Left | None | Left SFG | FCD iia | 1 (39) |
| 2 | Grid | F | Right | 12 | 36 | 24 | Left | Left | None | Left mesial SFG | Non-specific | 5 (54) |
| 3 | Grid | M | Right | 6 | 26 | 20 | Right | Bilateral | None | Right SFG/precentral | FCD iia | 4 (33) |
| 4 | Grid | F | Right | 10 | 28 | 18 | Right | Left | None | Right SFG | FCD iib | 1 (54) |
| 5 | Grid | F | Right | 8 | 23 | 15 | Left | Left | Left insula lesion | Left anterior frontal | Gliosis | 5 (27) |
| 6 | Grid | M | Right | 12 | 20 | 8 | Left | Left | None | Left pre/post central/SMA | Inoperable | N/A |
| 7 | Grid+depth | M | Right | 16 | 45 | 29 | Left | Left | None | Not localised | Inoperable | N/A |
| 8 | Grid+depth | M | Right | 2 | 21 | 19 | Right | Left | Dysplastic left precentral gyrus | Primary motor area | Inoperable | N/A |
| 9 | Grid+depth | M | Right | 38 | 48 | 10 | Left | Left | None | Left SMA | Inoperable | N/A |
| 10 | Grid+depth | M | Right | 10 | 23 | 13 | Right | N/a | None | Not localised | Inoperable | N/A |
| 11 | Grid+depth | M | Right | 6 | 25 | 19 | Left | Left | Left central gyrus signal change | Right hand sensory motor area | Inoperable | N/A |
| 12 | Grid+depth | F | Right | 11 | 21 | 10 | Left | Left | Cerebellar lesion | Left primary motor cortex | Inoperable | N/A |
| 13 | Grid+depth | M | Right | 8 | 18 | 10 | Left | Left | Left precentral gyrus tumour | Left paracentral/precentral | Inoperable | N/A |
| 14 | Grid+depth | M | Right | 6 | 38 | 32 | Right | Left | Dysplastic right SFG | Right SMA | Gliosis | 5 (42) |
| 15 | Grid+depth | M | Right | 29 | 41 | 12 | Left | Left | None | Left SMA | Non-specific | 5 (12) |
| 16 | Grid+depth | F | Right | 5 | 49 | 44 | Left | Left | Dysplastic left IFG | Left IFG | Inoperable | N/A |
| 17 | Grid+depth | M | Right | 12 | 32 | 20 | Right | Bilateral | Right parietal dysplasia | Right SPL | FCD iib | 1 (54) |
| 18 | Grid+depth | M | Right | 14 | 49 | 35 | Left | Left | Resected left MFG | Left SFG and MFG | DNET | 3 (19) |
| 19 | Depth | M | Right | 13 | 30 | 17 | Left | Left | Non lesional | Left anterior medial frontal | Non-specific | 5 (12) |
| 20 | Depth | M | Right | 15 | 29 | 14 | Left | Left | Non lesional | Left insula | Inoperable | N/A |
| 21 | Depth | M | Right | 25 | 46 | 21 | Right | Bilateral | Non lesional | Right orbito-/inferior frontal | Inoperable | N/A |
| 22 | Depth | M | Right | 12 | 30 | 18 | Right | Bilateral | Non lesional | Right mesiofrontal | Awaiting | N/A |
| 23 | Depth | M | Right | 10 | 33 | 23 | Right | Left | Non lesional | Right insular | Awaiting | N/A |
| 24 | Depth | F | Right | 13 | 46 | 33 | Left | Left | Left frontal dysplasia | Left middle frontal gyrus | Awaiting | N/A |
| 25 | Depth | M | Right | 7 | 32 | 25 | Left | Bilateral | Left HS | Left temporal | Awaiting | N/A |
| 26 | Depth | M | Equal | 1 | 37 | 36 | Left | Left | Left HS | Left hippocampus | HS | 1 (4) |
| 27 | Depth | M | Right | 23 | 27 | 4 | Right | Left | None | Right SFG and MFG | Awaiting | N/a |
| 28 | Depth | M | Right | 5 | 19 | 14 | Right | Unclear | None | Right SFG | FCD iib | 3 (13) |
| 29 | Depth | M | Right | 1 | 26 | 25 | Both | Left | None | Right SFG and MFG | Inoperable | N/A |
| 30 | Depth | F | Right | 6 | 22 | 16 | Left | N/a | None | Left anterior insula | Inoperable | N/A |
| 31 | Depth | M | Right | 7 | 41 | 34 | Left | Left | None | Left insula | Inoperable | N/A |
| 32 | Depth | M | Equal | 3 | 44 | 41 | Left | Right | Left frontal dysplasia | Left frontopolar/orbital | Awaiting | N/A |
| 33 | Depth | M | Right | 11 | 41 | 30 | Right | Left | Lesion SFG | Right SFG | Inoperable | 4 (15) |
| 34 | Depth | F | Right | 3 | 34 | 31 | Left | Left | Left frontal dysplasia | Frontal lobe | Inoperable | N/A |
| 35 | Depth | M | Right | 16 | 32 | 16 | Right | Left | None | Right orbitofrontal | Gliosis | 1 (36) |
| 36 | Depth | M | Right | 38 | 68 | 30 | Right | Left | None | Right SMA | Non-specific | 1 (60) |
| 37 | Depth | M | Left | 6 | 46 | 40 | Right | Left | None | Right orbitofrontal | Normal | 1 (6) |

**Supplementary Table T1. Patient clinical characteristics.** Summary of demographics and clinical characteristics of the included patients with stimulations in the medial frontal cortex, based on Trevisi et al. (2018). F = female; M = male; ICR = intracranial recording; FCD = focal cortical dysplasia; SMA = supplementary motor area; HS = hippocampal sclerosis; DNET = dysembryoplastic neuroepithelial tumour; SFG = superior frontal gyrus; SMG =superior medial gyrus; IFG = inferior frontal gyrus; SPL = superior parietal gyrus; ILAE = international league against epilepsy.

Supplementary Table T2.

Positive Motor (n = 153)

| | x | y | z | | x | y | z | | x | y | z |
|---|---|---|---|---|---|---|---|---|---|---|---|
| 1 | 1 | 1 | 42 | 52 | −9 | −19 | 69 | 103 | 25 | −18 | 70 |
| 2 | 10 | −23 | 51 | 53 | −8 | −13 | 57 | 104 | 29 | −18 | 71 |
| 3 | −18 | −16 | 54 | 54 | −5 | −6 | 44 | 105 | 6 | 1 | 47 |
| 4 | −39 | −3 | 39 | 55 | −4 | −3 | 45 | 106 | 7 | −23 | 58 |
| 5 | 1 | −30 | 55 | 56 | −4 | 2 | 43 | 107 | 10 | −24 | 62 |
| 6 | 2 | −10 | 40 | 57 | −4 | −2 | 41 | 108 | 11 | 3 | 52 |
| 7 | 3 | −1 | 47 | 58 | −3 | −7 | 54 | 109 | 14 | −23 | 66 |
| 8 | −8 | 14 | 48 | 59 | −3 | 0 | 47 | 110 | −22 | −6 | 55 |
| 9 | −5 | −1 | 60 | 60 | −3 | −5 | 49 | 111 | −21 | −23 | 66 |
| 10 | −2 | −2 | 64 | 61 | 0 | −30 | 58 | 112 | −16 | −23 | 64 |
| 11 | 23 | −13 | 32 | 62 | 0 | −6 | 59 | 113 | −12 | −9 | 52 |
| 12 | −2 | −12 | 52 | 63 | 0 | −1 | 56 | 114 | −7 | −11 | 51 |
| 13 | −1 | −9 | 57 | 64 | 2 | −1 | 52 | 115 | −22 | −4 | 65 |
| 14 | −1 | −31 | 43 | 65 | 4 | −9 | 52 | 116 | −17 | −3 | 63 |
| 15 | 0 | −6 | 62 | 66 | −2 | −13 | 69 | 117 | −12 | −2 | 62 |
| 16 | 3 | −4 | 67 | 67 | −2 | −8 | 67 | 118 | −9 | −1 | 61 |
| 17 | −1 | −21 | 37 | 68 | 2 | −15 | 65 | 119 | −3 | 2 | 57 |
| 18 | −1 | −29 | 49 | 69 | 2 | −9 | 63 | 120 | 0 | 4 | 55 |
| 19 | −1 | −34 | 59 | 70 | 4 | −16 | 59 | 121 | −33 | 2 | 50 |
| 20 | −1 | −40 | 68 | 71 | 4 | −10 | 57 | 122 | −22 | 1 | 49 |
| 21 | 0 | −22 | 59 | 72 | −32 | −24 | 30 | 123 | −17 | 1 | 49 |
| 22 | 0 | −21 | 70 | 73 | −2 | 9 | 35 | 124 | −11 | 1 | 50 |
| 23 | 0 | −17 | 58 | 74 | −2 | 12 | 41 | 125 | −6 | 1 | 50 |
| 24 | 0 | −7 | 61 | 75 | −2 | −6 | 50 | 126 | −1 | 1 | 50 |
| 25 | 1 | −45 | 74 | 76 | −2 | −3 | 56 | 127 | 5 | 1 | 49 |
| 26 | 5 | −39 | 62 | 77 | −2 | 0 | 67 | 128 | 10 | 1 | 48 |
| 27 | 19 | −20 | 66 | 78 | −2 | −10 | 55 | 129 | 0 | −21 | 64 |
| 28 | 0 | −8 | 62 | 79 | −2 | −7 | 66 | 130 | 1 | −32 | 63 |
| 29 | 0 | −13 | 64 | 80 | −2 | −15 | 58 | 131 | 2 | −27 | 61 |
| 30 | 1 | −3 | 61 | 81 | −2 | −13 | 64 | 132 | 4 | −18 | 56 |
| 31 | 2 | −4 | 56 | 82 | −1 | −12 | 69 | 133 | 4 | −29 | 56 |
| 32 | 2 | −8 | 56 | 83 | −22 | −15 | 60 | 134 | 4 | −36 | 53 |
| 33 | 2 | −14 | 58 | 84 | −5 | −22 | 47 | 135 | 5 | 13 | 27 |
| 34 | 3 | 0 | 48 | 85 | 0 | −8 | 48 | 136 | 5 | 16 | 31 |
| 35 | 3 | −6 | 49 | 86 | 0 | −22 | 45 | 137 | 5 | 5 | 34 |
| 36 | 3 | −10 | 51 | 87 | −37 | −9 | 55 | 138 | 5 | 9 | 39 |
| 37 | 3 | −15 | 53 | 88 | −31 | −9 | 54 | 139 | 5 | −19 | 51 |
| 38 | 4 | −7 | 44 | 89 | −25 | −9 | 51 | 140 | 5 | −23 | 54 |
| 39 | 4 | −12 | 45 | 90 | −20 | −9 | 48 | 141 | 5 | −31 | 51 |
| 40 | 4 | −17 | 47 | 91 | −13 | −9 | 46 | 142 | 6 | 9 | 31 |
| 41 | −7 | −12 | 56 | 92 | 2 | −14 | 57 | 143 | 6 | 1 | 38 |
| 42 | −4 | 11 | 39 | 93 | 6 | −15 | 61 | 144 | 6 | −21 | 47 |
| 43 | −9 | 9 | 55 | 94 | 29 | −16 | 71 | 145 | 6 | −26 | 50 |
| 44 | −7 | 17 | 39 | 95 | 7 | −14 | 55 | 146 | −1 | −28 | 62 |
| 45 | −33 | −18 | 51 | 96 | 7 | −19 | 59 | 147 | 3 | −37 | 67 |
| 46 | −28 | −31 | 51 | 97 | 9 | −6 | 47 | 148 | 0 | −12 | 45 |
| 47 | −14 | −18 | 65 | 98 | 12 | −5 | 51 | 149 | 0 | −12 | 57 |
| 48 | −13 | −23 | 66 | 99 | 12 | −19 | 62 | 150 | 0 | −12 | 62 |
| 49 | −13 | −33 | 66 | 100 | 16 | −19 | 64 | 151 | 4 | −11 | 68 |
| 50 | −13 | −36 | 65 | 101 | 20 | −18 | 68 | 152 | 9 | −11 | 69 |
| 51 | −10 | −24 | 68 | 102 | 25 | −18 | 69 | 153 | 14 | −11 | 70 |

## Negative Motor (n = 41)

|   | x | y | z |
|---|---|---|---|
| 1 | 1 | 1 | 42 |
| 2 | 2 | 8 | 55 |
| 3 | −8 | 14 | 48 |
| 4 | 5 | 2 | 66 |
| 5 | −3 | −14 | 47 |
| 6 | 1 | 2 | 59 |
| 7 | 2 | 1 | 54 |
| 8 | 3 | 0 | 48 |
| 9 | 3 | −6 | 49 |
| 10 | 4 | −7 | 44 |
| 11 | −7 | −12 | 56 |
| 12 | 8 | 42 | 44 |
| 13 | −8 | −13 | 57 |
| 14 | −5 | −6 | 44 |
| 15 | −4 | −2 | 41 |
| 16 | −3 | 0 | 47 |
| 17 | −3 | −5 | 49 |
| 18 | −2 | −8 | 58 |
| 19 | −1 | −3 | 55 |
| 20 | −6 | 18 | 46 |
| 21 | 2 | −1 | 52 |
| 22 | 4 | −9 | 52 |
| 23 | −32 | −24 | 30 |
| 24 | −19 | −2 | 53 |
| 25 | −4 | −7 | 48 |
| 26 | 0 | −8 | 48 |
| 27 | 1 | 17 | 48 |
| 28 | 5 | 19 | 52 |
| 29 | 8 | 12 | 37 |
| 30 | −8 | 13 | 54 |
| 31 | −3 | 12 | 53 |
| 32 | 21 | 5 | 64 |
| 33 | −27 | −5 | 55 |
| 34 | −22 | −6 | 55 |
| 35 | −17 | 9 | 63 |
| 36 | −17 | −8 | 53 |
| 37 | −12 | 8 | 59 |
| 38 | −9 | 7 | 56 |
| 39 | −6 | 5 | 52 |
| 40 | −2 | 4 | 48 |
| 41 | 0 | 3 | 42 |

## Sensory (n = 46)

|   | x | y | z |
|---|---|---|---|
| 1 | 3 | 6 | 46 |
| 2 | 10 | −23 | 51 |
| 3 | −39 | −3 | 39 |
| 4 | 2 | 8 | 55 |
| 5 | −8 | 14 | 48 |
| 6 | −2 | −2 | 64 |
| 7 | −4 | −16 | 41 |
| 8 | −2 | 14 | 53 |
| 9 | 1 | 18 | 57 |
| 10 | −7 | −12 | 56 |
| 11 | −13 | −40 | 65 |
| 12 | −3 | −9 | 52 |
| 13 | −5 | −9 | 41 |
| 14 | −5 | 0 | 37 |
| 15 | 2 | −3 | 61 |
| 16 | 4 | −5 | 55 |
| 17 | 4 | 1 | 53 |
| 18 | −2 | 5 | 40 |
| 19 | −2 | 0 | 44 |
| 20 | −2 | 2 | 50 |
| 21 | 4 | −9 | 38 |
| 22 | −2 | −9 | 41 |
| 23 | −8 | −9 | 44 |
| 24 | 5 | 13 | 49 |
| 25 | 7 | 14 | 52 |
| 26 | 11 | 14 | 56 |
| 27 | 4 | −9 | 38 |
| 28 | 7 | −7 | 43 |
| 29 | 14 | −3 | 55 |
| 30 | 16 | −2 | 58 |
| 31 | 29 | −18 | 71 |
| 32 | −7 | −11 | 51 |
| 33 | −28 | 2 | 49 |
| 34 | 5 | 13 | 27 |
| 35 | −2 | −32 | 69 |
| 36 | −1 | −26 | 66 |
| 37 | 3 | −22 | 59 |
| 38 | 5 | −31 | 51 |
| 39 | 4 | −29 | 56 |
| 40 | 2 | −27 | 61 |
| 41 | 4 | −36 | 53 |
| 42 | 3 | −34 | 58 |
| 42 | −1 | 1 | 40 |
| 44 | −1 | −6 | 45 |
| 45 | −1 | −20 | 58 |
| 46 | 0 | −12 | 51 |

## Speech (n = 46)

|   | x | y | z |
|---|---|---|---|
| 1 | −1 | 23 | 26 |
| 2 | 3 | 20 | 43 |
| 3 | 2 | 8 | 55 |
| 4 | −8 | 14 | 48 |
| 5 | 5 | 2 | 66 |
| 6 | 7 | 2 | 65 |
| 7 | 13 | −9 | 68 |
| 8 | 8 | 42 | 44 |
| 9 | 0 | 6 | 51 |
| 10 | −1 | −3 | 55 |
| 11 | −2 | −8 | 58 |
| 12 | −3 | −9 | 52 |
| 13 | −5 | −5 | 40 |
| 14 | −5 | 5 | 35 |
| 15 | −4 | −3 | 45 |
| 16 | −4 | 2 | 43 |
| 17 | −4 | −2 | 41 |
| 18 | −3 | 0 | 47 |
| 19 | −3 | −5 | 49 |
| 20 | 2 | −1 | 52 |
| 21 | −1 | 5 | 64 |
| 22 | −2 | −8 | 60 |
| 23 | −2 | −7 | 66 |
| 24 | −2 | −13 | 64 |
| 25 | −1 | −12 | 69 |
| 26 | −1 | −10 | 73 |
| 27 | 0 | −18 | 68 |
| 28 | 1 | −17 | 73 |
| 29 | 2 | −16 | 76 |
| 30 | −19 | −2 | 53 |
| 31 | −23 | −1 | 54 |
| 32 | −27 | 0 | 56 |
| 33 | −32 | 1.3 | 59 |
| 34 | 1 | 6 | 60 |
| 35 | −4 | 6 | 64 |
| 36 | −8 | 6 | 67 |
| 37 | −13 | 5 | 70 |
| 38 | −18 | 5 | 70 |
| 39 | 1 | 17 | 48 |
| 40 | 5 | 19 | 52 |
| 41 | 8 | 12 | 37 |
| 42 | −3 | 12 | 53 |
| 42 | −8 | 13 | 54 |
| 44 | 0 | 3 | 42 |
| 45 | −12 | 8 | 59 |
| 46 | 6 | 9 | 31 |

# No Response (n = 243)

|     | x   | y   | z   |
| --- | --- | --- | --- |
| 1   | 0   | 57  | 22  |
| 2   | 0   | 52  | 31  |
| 3   | 0   | 46  | 39  |
| 4   | 0   | 41  | 48  |
| 5   | 0   | 34  | 57  |
| 6   | −2  | 29  | 48  |
| 7   | −2  | 20  | 55  |
| 8   | −2  | 10  | 62  |
| 9   | −3  | 9   | 53  |
| 10  | 0   | 11  | 58  |
| 11  | 3   | 12  | 62  |
| 12  | 7   | 13  | 65  |
| 13  | −3  | 14  | 49  |
| 14  | 0   | 15  | 55  |
| 15  | 4   | 17  | 59  |
| 16  | 9   | 18  | 62  |
| 17  | −2  | 19  | 46  |
| 18  | 1   | 20  | 52  |
| 19  | 5   | 22  | 56  |
| 20  | 9   | 22  | 58  |
| 21  | −1  | 23  | 43  |
| 22  | 1   | 26  | 48  |
| 23  | 5   | 27  | 52  |
| 24  | 9   | 28  | 55  |
| 25  | −1  | 28  | 40  |
| 26  | 3   | 31  | 45  |
| 27  | 6   | 32  | 48  |
| 28  | 10  | 33  | 51  |
| 29  | 0   | 34  | 37  |
| 30  | 3   | 35  | 42  |
| 31  | 7   | 37  | 45  |
| 32  | 11  | 39  | 47  |
| 33  | 0   | 39  | 35  |
| 34  | 4   | 40  | 38  |
| 35  | 8   | 42  | 41  |
| 36  | 12  | 43  | 43  |
| 37  | 1   | 45  | 32  |
| 38  | 5   | 46  | 36  |
| 39  | 9   | 48  | 39  |
| 40  | 14  | −49 | 40  |
| 41  | 7   | 5   | 56  |
| 42  | −11 | 24  | 56  |
| 43  | 1   | 20  | 43  |
| 44  | 0   | 24  | 48  |
| 45  | −2  | 27  | 53  |
| 46  | 1   | 24  | 39  |
| 47  | 0   | 27  | 44  |
| 48  | −2  | 31  | 49  |
| 49  | 1   | 26  | 36  |
| 50  | 0   | 30  | 41  |
| 51  | −2  | 34  | 46  |
| 52  | 1   | 29  | 31  |
| 53  | 0   | 33  | 36  |
| 54  | −2  | 37  | 41  |
| 55  | 1   | 34  | 28  |
| 56  | 0   | 38  | 33  |
| 57  | −1  | 41  | 38  |
| 58  | 2   | 37  | 24  |
| 59  | 1   | 41  | 30  |
| 60  | 0   | 45  | 34  |
| 61  | 2   | 40  | 21  |
| 62  | 1   | 45  | 26  |
| 63  | 0   | 48  | 30  |
| 64  | 2   | 44  | 17  |
| 65  | 2   | 49  | 22  |
| 66  | 1   | 52  | 27  |
| 67  | −5  | 30  | 48  |
| 68  | −1  | −15 | 69  |
| 69  | 2   | −13 | 69  |
| 70  | 6   | −8  | 68  |
| 71  | 7   | −2  | 70  |
| 72  | −5  | −19 | 35  |
| 73  | −2  | 16  | 44  |
| 74  | −2  | 19  | 49  |
| 75  | 2   | 22  | 53  |
| 76  | 4   | 25  | 57  |
| 77  | −3  | 11  | 48  |
| 78  | 4   | 20  | 60  |
| 79  | 0   | −16 | 69  |
| 80  | 0   | −13 | 67  |
| 81  | 0   | −13 | 56  |
| 82  | 0   | −6  | 66  |
| 83  | 4   | −2  | 42  |
| 84  | −60 | −62 | 29  |
| 85  | −55 | −59 | 37  |
| 86  | −50 | −55 | 44  |
| 87  | −46 | −53 | 52  |
| 88  | −40 | −50 | 60  |
| 89  | −33 | −46 | 67  |
| 90  | −24 | −43 | 72  |
| 91  | −14 | −39 | 72  |
| 92  | −63 | −52 | 27  |
| 93  | −60 | −47 | 35  |
| 94  | −55 | −44 | 45  |
| 95  | −51 | −42 | 54  |
| 96  | −44 | −38 | 61  |
| 97  | −36 | −35 | 67  |
| 98  | −28 | −32 | 72  |
| 99  | −18 | −28 | 73  |
| 100 | −65 | −40 | 23  |
| 101 | −64 | −37 | 34  |
| 102 | −59 | −33 | 45  |
| 103 | −54 | −30 | 53  |
| 104 | −47 | −27 | 61  |
| 105 | −38 | −23 | 66  |
| 106 | −29 | −20 | 70  |
| 107 | −21 | −16 | 71  |
| 108 | −67 | −32 | 19  |
| 109 | −63 | −28 | 30  |
| 110 | −60 | −23 | 41  |
| 111 | −55 | −19 | 50  |
| 112 | −49 | −16 | 59  |
| 113 | −41 | −13 | 63  |
| 114 | −31 | −10 | 66  |
| 115 | −22 | −7  | 68  |
| 116 | −2  | 39  | 42  |
| 117 | −9  | 40  | 22  |
| 118 | −11 | 46  | 39  |
| 119 | 8   | 44  | 28  |
| 120 | −15 | −14 | 64  |
| 121 | −1  | 52  | 25  |
| 122 | −5  | 47  | 15  |
| 123 | −1  | 47  | 28  |
| 124 | −5  | 43  | 18  |
| 125 | 0   | 41  | 34  |
| 126 | −2  | 39  | 29  |
| 127 | −5  | 37  | 25  |
| 128 | −5  | 34  | 19  |
| 129 | −1  | 27  | 44  |
| 130 | −3  | 25  | 38  |
| 131 | 0   | 22  | 46  |
| 132 | −5  | 19  | 34  |
| 133 | 0   | 17  | 47  |
| 134 | −5  | 14  | 36  |
| 135 | −3  | 9   | 37  |
| 136 | −4  | 3   | 40  |
| 137 | −1  | 1   | 53  |
| 138 | −5  | −11 | 46  |
| 139 | −5  | −8  | 47  |
| 140 | −4  | 6   | 41  |
| 141 | −2  | 3   | 49  |
| 142 | 1   | 4   | 54  |
| 143 | −3  | 4   | 45  |
| 144 | −2  | −1  | 65  |
| 145 | 4   | 11  | 48  |
| 146 | 5   | 11  | 41  |
| 147 | 5   | 13  | 35  |
| 148 | 3   | 17  | 50  |
| 149 | 3   | 18  | 43  |
| 150 | 4   | 19  | 36  |
| 151 | 2   | 23  | 51  |
| 152 | 2   | 23  | 44  |
| 153 | 3   | 24  | 37  |

|     | x   | y   | z  |
|-----|-----|-----|----|
| 154 | 0   | 28  | 51 |
| 155 | 0   | 28  | 44 |
| 156 | 1   | 29  | 38 |
| 157 | −6  | 26  | 50 |
| 158 | −4  | 39  | 28 |
| 159 | 0   | 46  | 38 |
| 160 | −4  | 43  | 25 |
| 161 | 0   | 50  | 34 |
| 162 | −4  | 46  | 21 |
| 163 | 0   | 52  | 31 |
| 164 | −4  | 50  | 18 |
| 165 | 0   | 57  | 27 |
| 166 | 2   | 67  | 5  |
| 167 | 3   | 59  | 26 |
| 168 | 1   | 63  | 3  |
| 169 | 1   | 61  | 15 |
| 170 | 1   | 56  | 26 |
| 171 | 1   | 52  | 39 |
| 172 | 1   | 46  | 53 |
| 173 | 1   | 39  | 62 |
| 174 | 2   | 13  | 32 |
| 175 | −2  | 17  | 36 |
| 176 | −2  | 19  | 43 |
| 177 | −2  | 21  | 50 |
| 178 | −2  | 14  | 47 |
| 179 | −2  | 17  | 54 |
| 180 | −2  | 7   | 46 |
| 181 | −2  | 8   | 52 |
| 182 | −2  | 12  | 58 |
| 183 | −2  | 5   | 57 |
| 184 | −2  | 6   | 62 |
| 185 | −2  | −1  | 61 |
| 186 | −1  | −5  | 70 |
| 187 | −1  | −20 | 62 |
| 188 | 1   | 20  | 33 |
| 189 | −1  | 19  | 39 |
| 190 | −14 | 19  | 67 |
| 191 | 17  | 22  | 64 |
| 192 | 2   | 12  | 36 |
| 193 | 33  | 15  | 45 |
| 194 | 13  | 16  | 60 |
| 195 | 16  | 17  | 64 |
| 196 | −12 | 13  | 55 |
| 197 | 14  | 4   | 56 |
| 198 | 18  | 5   | 60 |
| 199 | −21 | 10  | 65 |
| 200 | −24 | −22 | 69 |
| 201 | 0   | 32  | 43 |
| 202 | −4  | 35  | 46 |
| 203 | −7  | 39  | 47 |

|     | x   | y   | z  |
|-----|-----|-----|----|
| 204 | −2  | 51  | 16 |
| 205 | −2  | 48  | 27 |
| 206 | 0   | 42  | 37 |
| 207 | 3   | 38  | 46 |
| 208 | 5   | 33  | 54 |
| 209 | 9   | 28  | 61 |
| 210 | 4   | 20  | 35 |
| 211 | 5   | 12  | 36 |
| 212 | 4   | 17  | 40 |
| 213 | 4   | 13  | 43 |
| 214 | 6   | 5   | 42 |
| 215 | 5   | 9   | 46 |
| 216 | 4   | 40  | 6  |
| 217 | 4   | 40  | 19 |
| 218 | 4   | 40  | 31 |
| 219 | 4   | 35  | 42 |
| 220 | 0   | 31  | 50 |
| 221 | −1  | −14 | 52 |
| 222 | 0   | −12 | 39 |
| 223 | −2  | −37 | 36 |
| 224 | −1  | −47 | 31 |
| 225 | −2  | −40 | 41 |
| 226 | −2  | −49 | 36 |
| 227 | −2  | −41 | 46 |
| 228 | −2  | −51 | 40 |
| 229 | −2  | −43 | 52 |
| 230 | −2  | −53 | 45 |
| 231 | −1  | −45 | 55 |
| 232 | −1  | −54 | 50 |
| 233 | 2   | −47 | 60 |
| 234 | 1   | −56 | 54 |
| 235 | 5   | −48 | 64 |
| 236 | 5   | −56 | 58 |
| 237 | 10  | −48 | 65 |
| 238 | 9   | −56 | 63 |
| 239 | 3   | 51  | 7  |
| 240 | 3   | 47  | 18 |
| 241 | 3   | 43  | 30 |
| 242 | 3   | 39  | 41 |
| 243 | −1  | 35  | 54 |

Supplementary Table T3.

| Behaviour | t-statistic | p value (FWE corrected) | MNI Coordinates | | | Anatomical Location |
|---|---|---|---|---|---|---|
| | | | x | y | z | |
| Positive Motor | 7.94 | <0.000 | 4 | −10 | 60 | Supplementary motor area |
| | 3.92 | 0.049 | 12 | −15 | 66 | Precentral gyrus (medial segment) |
| Negative Motor | 7.15 | <0.000 | 0 | −4 | 48 | Supplementary motor area |
| Sensory | 5.91 | <0.000 | −2 | −6 | 34 | Middle cingulate gyrus |
| | 5.21 | <0.000 | 9 | −33 | 56 | Precentral gyrus (medial segment) |
| Speech | 5.36 | <0.000 | −10 | 6 | 66 | Superior frontal gyrus |
| | 5.25 | <0.000 | −3 | −14 | 76 | Superior frontal gyrus |
| | 3.96 | 0.042 | −22 | 3 | 57 | Pre-supplementary motor area |
| No Response | 5.21 | <0.000 | 2 | 18 | 38 | Middle cingulate gyrus |
| | 4.36 | 0.010 | −8 | 32 | 44 | Superior frontal gyrus (medial segment) |

**Supplementary Table T3. Statistical results of local disruptive mapping.** For each of the behavioural effects, the t-statistics, cluster size, and *p* values (FWE corrected) and MNI coordinates of the peak-level voxels are provided, as well as the nearest grey-matter location.

Supplementary Table T4.

| Behaviour | t-statistic | p value (FWE corrected) | MNI Coordinates | | | Anatomical Location |
|---|---|---|---|---|---|---|
| | | | x | y | z | |
| Positive Motor | 10.17 | <0.000 | 0 | −74 | −42 | Fronto-parietal cerebellar network |
| | 8.74 | <0.000 | 4 | −32 | 66 | Precentral gyrus (medial segment) |
| | 7.74 | <0.000 | 14 | −88 | 30 | Superior occipital gyrus |
| | 7.62 | <0.000 | 0 | 0 | 2 | Thalamus |
| | 7.43 | <0.000 | −16 | −50 | 60 | Superior parietal lobule |
| | 6.13 | <0.000 | −18 | −86 | 28 | Superior occipital gyrus |
| | 5.77 | <0.000 | 32 | −22 | 16 | Posterior insula |
| | 5.64 | 0.001 | −18 | −26 | −4 | Thalamus |
| | 5.44 | 0.001 | 20 | −16 | 20 | Caudate |
| | 5.33 | 0.002 | −16 | −2 | 60 | Superior frontal gyrus |
| | 5.00 | 0.011 | 26 | −4 | 0 | Pallidum |
| | 4.90 | 0.016 | 42 | −70 | −36 | Sensory-motor foot cerebellar network |
| | 4.88 | 0.017 | −30 | −64 | −60 | Sensory-motor hand cerebellar network |
| | 4.63 | 0.048 | 48 | −8 | 14 | Central operculum |
| Negative Motor | 6.15 | <0.000 | −8 | −60 | 64 | Superior parietal lobule |
| | 6.00 | <0.000 | 0 | −10 | 52 | Supplementary motor area |
| | 5.92 | <0.000 | 0 | 18 | 50 | Supplementary motor area |
| | 5.71 | <0.000 | −10 | −76 | −36 | Dorsal attention cerebellar network |
| | 5.57 | 0.001 | −14 | 8 | 50 | Middle frontal gyrus |
| | 5.45 | 0.001 | −16 | 0 | 72 | Superior frontal gyrus |
| | 5.21 | 0.004 | −20 | −64 | −46 | Fronto-parietal cerebellar network |
| | 5.16 | 0.005 | 20 | −12 | 66 | Precentral gyrus |
| | 5.12 | 0.006 | 2 | −72 | −18 | Fronto-parietal cerebellar network |
| | 4.94 | 0.013 | 0 | −34 | 68 | Precentral gyrus (medial segment) |
| | 4.90 | 0.016 | 16 | −2 | 52 | Superior frontal gyrus |
| | 4.88 | 0.017 | 4 | −74 | 46 | Precuneus |
| Sensory | 5.48 | 0.001 | −4 | −18 | 68 | Precentral gyrus (medial segment) |
| | 5.40 | 0.002 | −2 | −40 | −70 | Medulla |
| | 4.96 | 0.012 | 0 | −70 | −22 | Fronto-parietal cerebellar network |
| | 4.87 | 0.017 | −6 | −2 | 0 | Thalamus |
| | 4.80 | 0.023 | 0 | −6 | 46 | Middle cingulate gyrus |
| No Response | 9.05 | <0.000 | 0 | 36 | 42 | Superior frontal gyrus (medial segment) |
| | 8.26 | <0.000 | 22 | −84 | −40 | Dorsal attention cerebellar network |
| | 7.60 | <0.000 | 0 | −54 | 14 | Precuneus |
| | 6.49 | <0.000 | 20 | −60 | 30 | Precuneus |
| | 5.60 | 0.001 | −36 | −76 | 38 | Angular gyrus |
| | 5.54 | 0.001 | 24 | 12 | 44 | Middle frontal gyrus |
| | 4.89 | 0.017 | −14 | 22 | −24 | Medial orbital gyrus |

**Supplementary Table T4. Statistical results of connective disruptive mapping.** For each of the behavioural effects, the t-statistics, cluster size, and *p* values (FWE-corrected) and MNI coordinates of the peak-level voxels are provided, as well as the nearest grey-matter location.

Supplementary Figure F1.

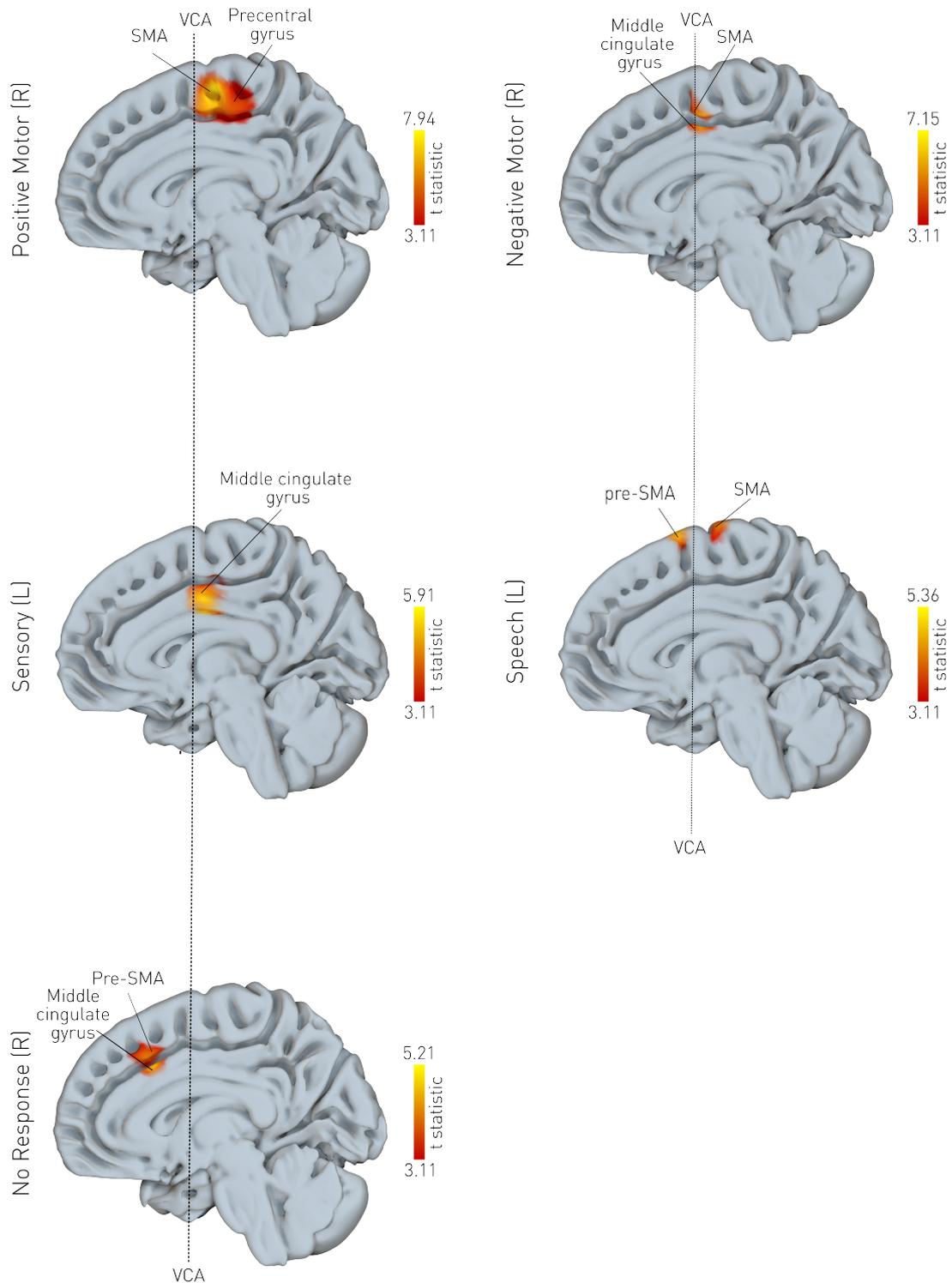

**Supplementary Figure F1. Local disruptive mapping of behaviour.** For each MNI voxel, a planned t-contrast was performed. Only voxels surviving the $p < 0.001$ uncorrected threshold are shown, overlaid on the mid-sagittal plane, where higher t-statistics (brighter colour) represent a stronger association between the electrode density value and the observed behaviour. R = right; L = left; (pre-)SMA = supplementary motor area.

Supplementary Figure F2.

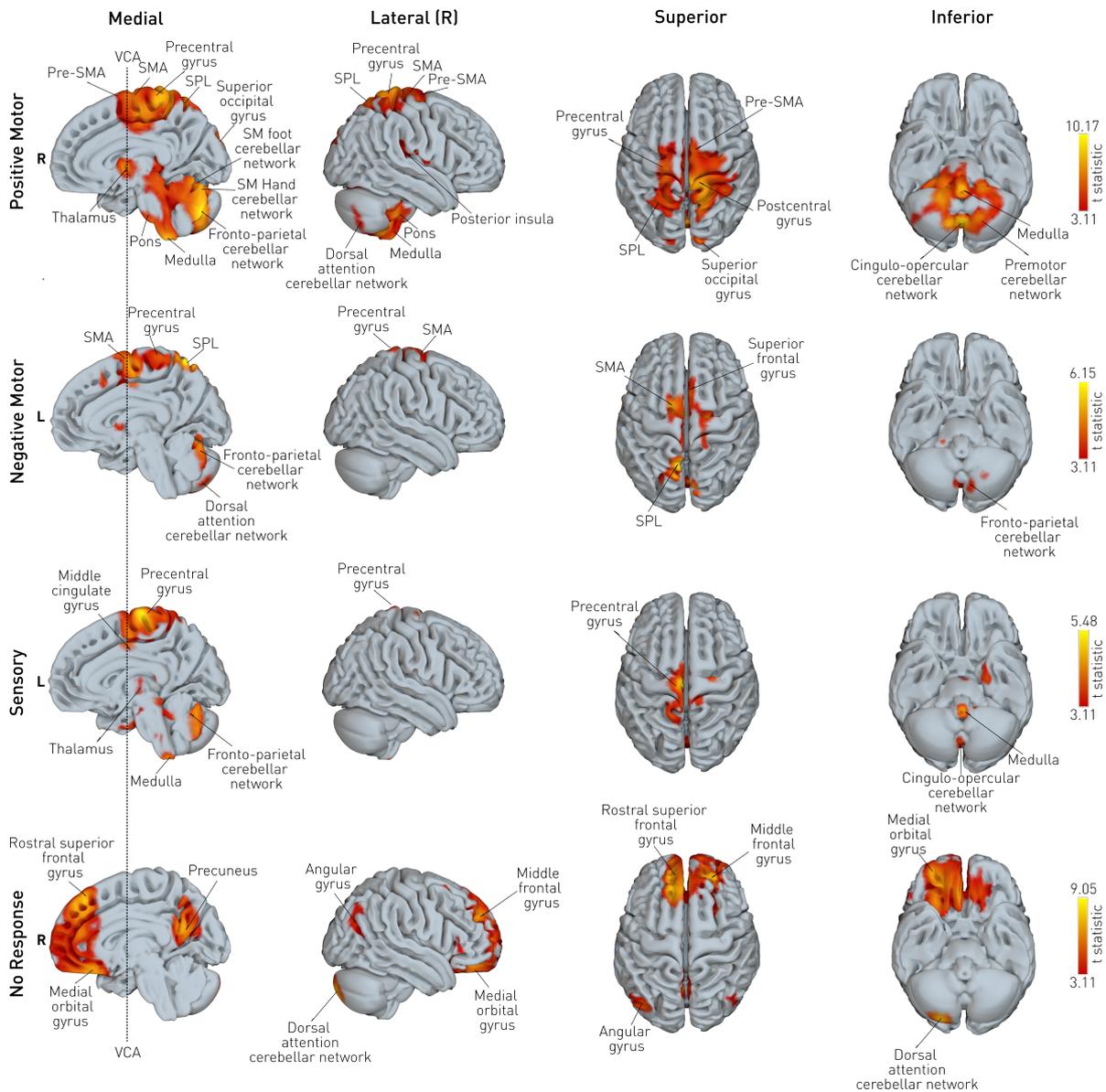

**Supplementary Figure F2. Connective disruptive maps of behaviour.** For each MNI voxel, a planned t-contrast was performed. Only voxels surviving the $p < 0.001$ uncorrected threshold are shown, overlaid on the mid-sagittal, lateral, superior, and inferior planes, where higher t-statistics (brighter colour) represent a stronger association between the connectivity value and the observed behaviour. Cerebellar subregions are labelled with reference to a priori known cortical network associations (Marek et al., 2018). R = right; L = left; SMA = supplementary motor area; SPL = superior parietal lobule, SM = sensorimotor).

## Supplementary Figure F3.

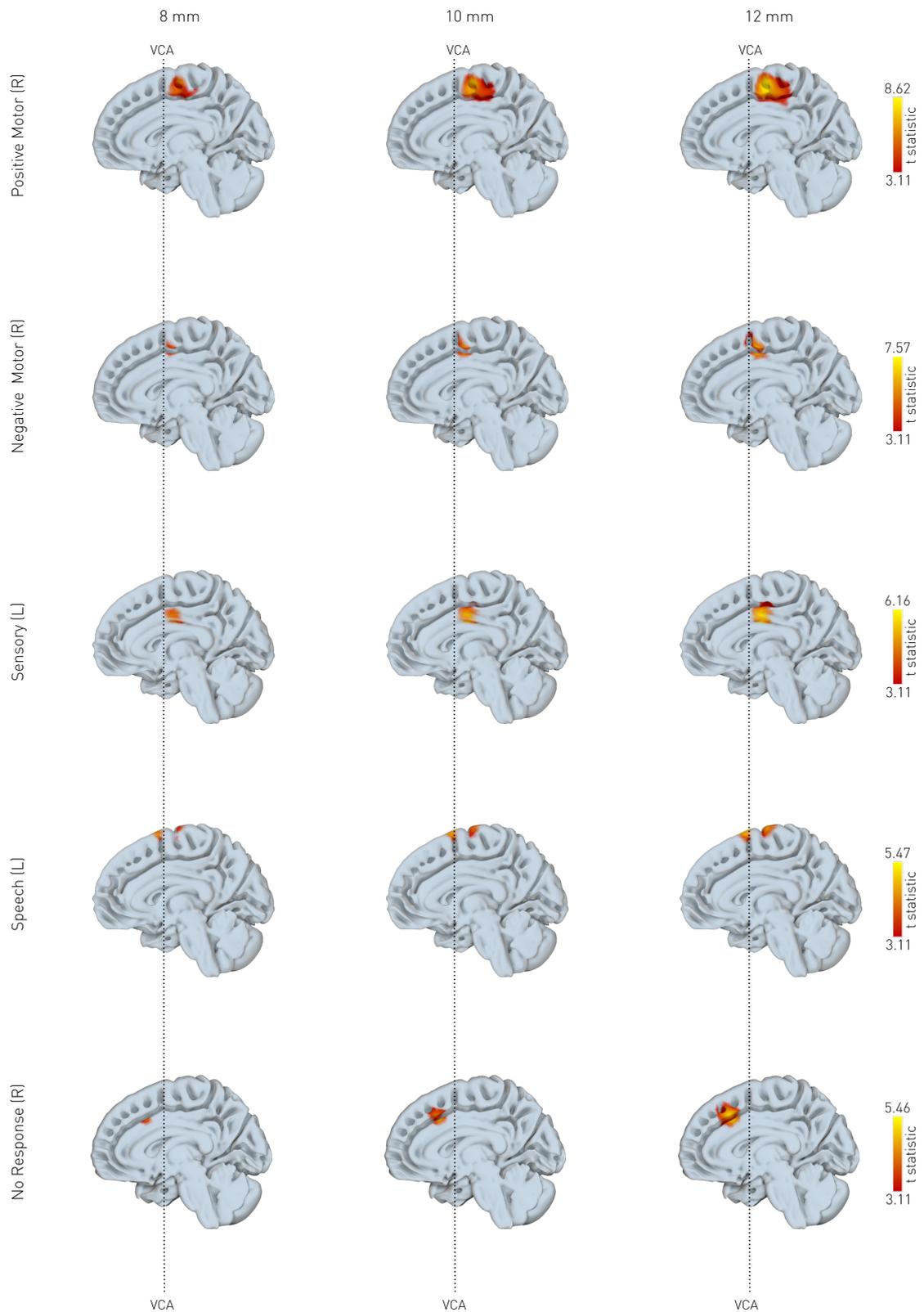

**Supplementary Figure F3. Sensitivity Analysis of Local disruptive mapping of behaviour.** For each MNI voxel, a planned t-contrast was performed. Only voxels surviving the $p < 0.001$ uncorrected threshold are shown, overlaid on the mid-sagittal plane, where higher t-statistics (brighter colour) represent a stronger association between the electrode density value and the observed behaviour. R = right; L = left; Analysis was replicated using different kernel sizes (8, 10, 12 mm isotropic) showing the results are not dependent on the chosen kernel size of 10 mm.